\documentclass[]{aa}

\usepackage[normalem]{ulem}
\usepackage{graphicx}
\usepackage{footmisc}
\usepackage{txfonts}
\usepackage{natbib}
\bibpunct{(}{)}{;}{a}{}{,}
\usepackage{scrextend}			
\usepackage{tikz}
\usepackage{subcaption}
\usepackage{multirow}
\usepackage{hyperref}
\usepackage{amsmath}  
\usepackage{xfrac}   
\usepackage{amssymb}	
\usepackage{graphicx}
\usepackage{amsmath}  
\usepackage{xfrac}   
\usepackage{amssymb}	
\usepackage{ dsfont }
\usepackage{txfonts}
\usepackage{mwe,tikz}\usepackage[percent]{overpic}
\begin{document}

  \title{A unified accretion-ejection paradigm for black hole X-ray binaries}

  \subtitle{III. Spectral signatures of hybrid disk configurations}

  \author{G.~Marcel\inst{1}
          \and
          J.~Ferreira\inst{1} 
          \and 
          P-O.~Petrucci\inst{1}
          \and
          R.~Belmont\inst{2}
          \and
          J.~Malzac\inst{2}
          \and
          M.~Clavel\inst{1}
          \and
          G.~Henri\inst{1}
          \and
          M.~Coriat\inst{2}
          \and
          S.~Corbel\inst{3}
          \and
          J.~Rodriguez\inst{3}
          \and
          A.~Loh\inst{3}
          \and
          S.~Chakravorty\inst{4}
          }
         
   \institute{Univ. Grenoble Alpes, CNRS, IPAG, 38000 Grenoble, France\\
              \email{gregoire.marcel@univ-grenoble-alpes.fr or gregoiremarcel26@gmail.com}
              \and
              IRAP, Université de Toulouse, CNRS, CNES, UPS, Toulouse, France
              \and
              Laboratoire AIM (CEA/IRFU - CNRS/INSU - Universite Paris Diderot), CEA DSM/IRFU/SAp, F-91191 Gif-sur-Yvette, France
              \and 
              Department of Physics, Indian Institute of Science, Bangalore 560012, India
             }

   \date{Received March 29, 2018; accepted May 31, 2018}

   \abstract
   {It has been suggested that the cycles of activity of X-ray Binaries (XrB) are triggered by a switch in the dominant disk torque responsible for accretion (paper I). As the disk accretion rate increases, the disk innermost regions would thus change from a jet-emitting disk (JED) to a standard accretion disk (SAD).}
   {While JEDs have been proven to successfully reproduce X-ray binary hard states (paper II), the existence of an outer cold SAD introduces an extra non local cooling term. We investigate the thermal structure and associated spectra of such a hybrid disk configuration.}
   {We use the two-temperature plasma code elaborated in paper II, allowing to compute outside-in the disk local thermal equilibrium with self-consistent advection and optically thin-to-thick transitions, in both radiation and gas supported regimes. The non-local inverse Compton cooling introduced by the external soft photons is computed by the \textsc{BELM} code.}  
   {This additional cooling term has a profound influence on JED solutions, allowing a smooth temperature transition from the outer SAD to the inner JED. We explore the full parameter space in disk accretion rate and transition radius, and show that the whole domain in X-ray luminosities and hardness ratios covered by standard XrB cycles is well reproduced by such hybrid disk configurations. Precisely, a reasonable combination of these parameters allows to reproduce the $3-200$~keV spectra of each of five canonical XrB states. Along with these X-ray signatures, JED-SAD configurations also naturally account for the radio emission whenever it is observed.}
   {By varying only the radial transition radius and the accretion rate, hybrid disk configurations combining an inner JED and an outer SAD are able to reproduce simultaneously the X-ray spectral states and radio emission of X-ray binaries during their outburst. Adjusting these two parameters, it is then possible to reproduce a full cycle. This will be shown in a forthcoming paper (paper IV).}


   \maketitle
%

\section{Introduction}

X-ray Binaries (hereafter XrBs) display cycles of strong activity, where their luminosity increases by several orders of magnitude and their spectral shape changes drastically on long timescales, before decreasing back to quiescence. We call this entire phase an outburst. At the beginning of such an outburst, they are in the so-called hard state where their X-ray spectra are dominated by a power-law component with a hard photon index $\Gamma<2$ \citep{Remillard06}. During this state, they also show flat or slightly inverted radio spectra \citep[e.g.,][]{CorbelFender02}, interpreted as self-absorbed synchrotron emission from collimated, mildly relativistic jets \citep{BK79}. At some point, the X-ray spectrum of these objects undergoes a smooth transition from this power-law dominated spectral shape, to a dominant blackbody of temperature $\sim~1$~keV only. In addition, steady radio emission disappears, suggesting a quenching of the jets \citep{Corbel04, Fender04, Fender09}. X-ray binaries remain in this soft state until a decline in luminosity makes them transit back to a hard state at the end of the outburst, along with reappearance of the jets. This surprising behavior has been observed multiple times in the past decades, and in dozens of different objects, where some have even undergone multiple outbursts \citep[see][for a global overview]{Dunn10}. What is even more striking in these outbursts, is that they seem to be very similar in different objects, while being different from one outburst to another in the same object! \\

A general scenario has been proposed by \citet{Esin97}, where changes in the accretion flow geometry provokes the spectral variations. In this view, the interplay between two different accretion flows is responsible for the spectral changes in the disk: in the outer parts, a cold standard accretion disk \citep[SAD, ][]{SS73} extends down to a given truncation (or transition) radius where an advection dominated hot flow\footnote{The inner hot flow is often referred to as a "hot corona". However, this designation remains ambiguous and we choose not to utilize it.} takes place \citep{Ichimaru77, Rees82}. The inner hot flow is expected to be responsible for the power-law component, while the outer cold flow produces the blackbody radiation. While the presence of a SAD in the outer regions of the disk is highly accepted to date \citep{Done07}, the physical properties of the advection dominated inner flow remain an open question.

Between Slim disks \citep{Abramowicz88}, ADAFs \citep{Narayan94}, ADIOS \citep{BB99}, LHAFs \citep{Yuan01} or more peculiar models \citep[e.g.,][]{Meyer00, Lasota01}, no satisfactory explanation has been provided so far \citep{Yuan14}. A discussion about the major models and their current state can be found in \citet{paperII}. Many questions remain open but in this article we focus on: (1) reproducing the X-ray spectral shape of all the generic spectral states, (2) explaining the correlated accretion-ejection processes through their observables, i.e. radio and X-rays fluxes. \\

In this work we consider an accretion flow extended down to the inner-most stable circular orbit, and thread by a large scale vertical magnetic field $B_z$.
It is well known that matter can only accrete by transferring away its angular momentum. This can be achieved by few physical mechanisms, namely internal turbulent ("viscous") torques and magnetic torques from an outflow:
\begin{itemize}
\item When accretion is mostly due to internal (turbulent) viscosity, angular momentum is transported radially. This produces an optically thick and geometrically thin accretion disk, a SAD, which is observed as a cold multicolor-disk blackbody. The production of winds by SAD is still a debated question, but few simulations and observations have shown the possible existence of winds from standard disks (see discussion in section~\ref{sec:innerSAD}). However, these winds cannot explain the powerful jets associated with XrBs so that the SAD suits perfectly for the jet-less thermal states in XrBs, i.e. soft states.
\item Alternatively, self-confined super-Alfv\'enic jets can also provide a feedback torque on the disk, carrying away vertically both energy and angular momentum. This accretion mode, referred to as jet-emitting disks \citep[JED, ][and subsequent work]{Ferreira95}, presents a supersonic accretion speed. Therefore, for the same accretion rate, it has a much smaller density than the SAD, leading to optically thin and geometrically thick disks. Disks accreting under this JED mode are thus good candidates to explain power-law dominated and jetted states in XrBs, i.e. hard states.
\end{itemize}

\noindent The magnetic field strength is characterized by the mid-plane magnetization $\mu(r)= B_z^2/\mu_o P_{tot}$, where $P_{tot}$ is the total pressure, the sum of the kinetic plasma pressure and the radiation pressure. At large magnetization, the SAD can no longer be maintained as magneto-centrifugally driven jets are launched: a JED arises \citep{Ferreira95,Ferreira97}. Full MHD calculations of JEDs have shown that the transition occurs around $\mu \sim 0.1$ \citep{Casse00a,Casse00b,Lesur13}. \\

A global scenario based on these possible dynamical transitions in accretion modes has then been proposed to explain XrB cycles (\citealt{Ferreira06}, hereafter paper I; \citealt{Petrucci08}). What would be the causes of the evolution of the disk magnetization distribution $\mu(r)$ is still a highly debated question. The main uncertainty comes from the interplay between the magnetic field advection and diffusion in turbulent accretion disks, geometrically thin or thick and with or without jets. Modern global 3D MHD simulations do show that large scale magnetic fields are indeed advected \citep{Avara16, ZhuStone18}, but these simulations are always done on quite short time scales, up to few seconds, and it is hard to scale them to the duration of XrB cycles, typically lasting over several month. In this paper, we assume that cycles result from transitions in accretion modes and focus on their observational consequences. \\

\citet{Petrucci10} computed the thermal states of a pure JED solutions and successfully reproduced the spectral emission, jet power and jet velocity during hard states of Cygnus X-1. However, their calculations were done assuming a one temperature (1T) plasma, but the necessity of a (2T) plasma seem inevitable to cover the large variation of accretion rate expected during an entire outburst \citep{Yuan14}.

\citet{paperII}, hereafter paper II, extended this work by developing a two-temperature (2T) plasma code that computes the disk local thermal equilibrium, including advection of energy and addressing optically thin-to-thick transitions in both radiation and gas supported regimes.
For a range in radius and accretion rates, they showed that JEDs exhibit three thermal equilibria, one thermally unstable and two stable ones. Only the stable equilibria are of physical importance \citep{Frank92}. One solution consists of a cold plasma, leading to an optically thick and geometrically thin disk, whereas the second solution describes a hot plasma, leading to an optically thin and geometrically thick disk. Due to the existence of these two thermally stable solutions a hysteresis cycle is naturally obtained. But large outbursting cycles, such as those exhibited by GX 339-4, cannot be reproduced (paper II). Nevertheless, JEDs have the striking property of being able to reproduce very well hard states spectral shape, all the way up to very luminous hard states $L > 30 \% L_{Edd}$.\\

However, SAD-JED local transitions are expected to occur locally on dynamical time scales, typically $\sim 1$~ms Kepler orbital time at $10~R_g$, whereas hard-to-soft transitions involve time scales of days or even weeks. This implies that, at any given time, the disk must be in some hybrid configuration with some regions emitting jets, while others do not. It is expected that jets, namely magnetocentrifugally-driven flows \citep{BP82}, are only launched from the disk innermost regions. This translates into a hybrid configuration where an inner JED is established from the last stable orbit $R_{in}$ until an unknown transition radius $R_J$, and then surrounded by an outer SAD until $R_{out}$. The exact location of the transition $R_J$ depends on the global response of the magnetic field $B_z$ to accretion rate evolution at the outer edge $\dot{M}_{out}$, two unknowns. Hence $R_J$ is treated here as a free parameter of the model. Such radial transition between two flows has already been studied in the context of non-magnetized accretion flows, advocating for mechanism such as evaporation or turbulent diffusion at the origin of the transition \citep[see, e.g.,][]{Meyer94, Honma96}. We study however configurations where the magnetization $\mu$ is large (near equipartition) and uniform in the JED region, and drops at the transition radius $R_J$, organizing the two-flow structure. Although the main properties of isolated JED and SAD are well understood, hybrid configurations imply mutual interactions that need to be described. For instance, part of cold radiation emitted from the SAD region must be intercepted by the geometrically thick JED and provides an additional cooling term that might change its general properties. \\

In this paper, we explore the observational signatures of disk configurations with an inner JED and an outer SAD. Section~\ref{sec2} describes this hybrid configuration, including interactions between the two regions, and explores some of its dynamical consequences. Section~\ref{sec3} presents the procedure followed to simulate and fit synthetic X-ray data from our theoretical spectra as well as to estimate the jet radio emission. Section~\ref{sec4} is devoted to the exploration of the parameter space, by varying the disk accretion rate $\dot{M}_{in}$ and transition radius $R_J$. Playing with these two parameters allows to completely cover the disk fraction luminosity diagram \citep[hereafter DFLD, ][]{Kording06}. As an illustrative example, we apply our model and reproduce canonical states of GX 339-4, both in X-ray spectral shape and radio fluxes. We end with concluding remarks in section~\ref{sec:Ccl}.

\section{Hybrid disk configuration: internal JED and external SAD} \label{sec2}

\subsection{General properties}   
\label{sec:GeneralPpts}

As introduced in paper II, we consider an axisymmetric accretion disk orbiting a black hole of mass $M$. For simplicity, the disk is assumed to be in global steady-state so that any radial variation of the disk accretion rate $\dot{M}(R)$ is only due to mass loss in outflows. We define $H(R)$ the half-height of the disk, $\varepsilon(R) = H/R$ its aspect ratio, $\dot{M}(R) = - 4 \pi R u_R \Sigma$ the local disk accretion rate, $u_R$ the radial (accretion) velocity and $\Sigma = \rho_0 H$ the vertical column density with $\rho_0$ the mid plane density. Throughout the paper, calculations are done within the Newtonian approximation. Moreover, and for the sake of simplicity, the disk is assumed to be always quasi-Keplerian with a local angular velocity $\Omega \simeq \Omega_K = \sqrt[]{G M R^{-3}}$, where $G$ is the gravitational constant. \\

The disk is assumed to be thread by a large scale vertical magnetic field $B_z(R)$. We assume that such a field is the result of field advection and diffusion and we neglect thereby any field amplification by dynamo.
%
Clearly, the existence of cycles shows that some evolution is ongoing within the disk. However, the timescales involved (days to months) are always much longer than accretion timescales inferred from the X-ray emitting regions. Thus, as for any other disk quantity, the local magnetic field is assumed to be stationary on dynamical time scales (Keplerian orbital time). \\

\begin{figure}[h!]
   \centering
   \includegraphics[width=\columnwidth]{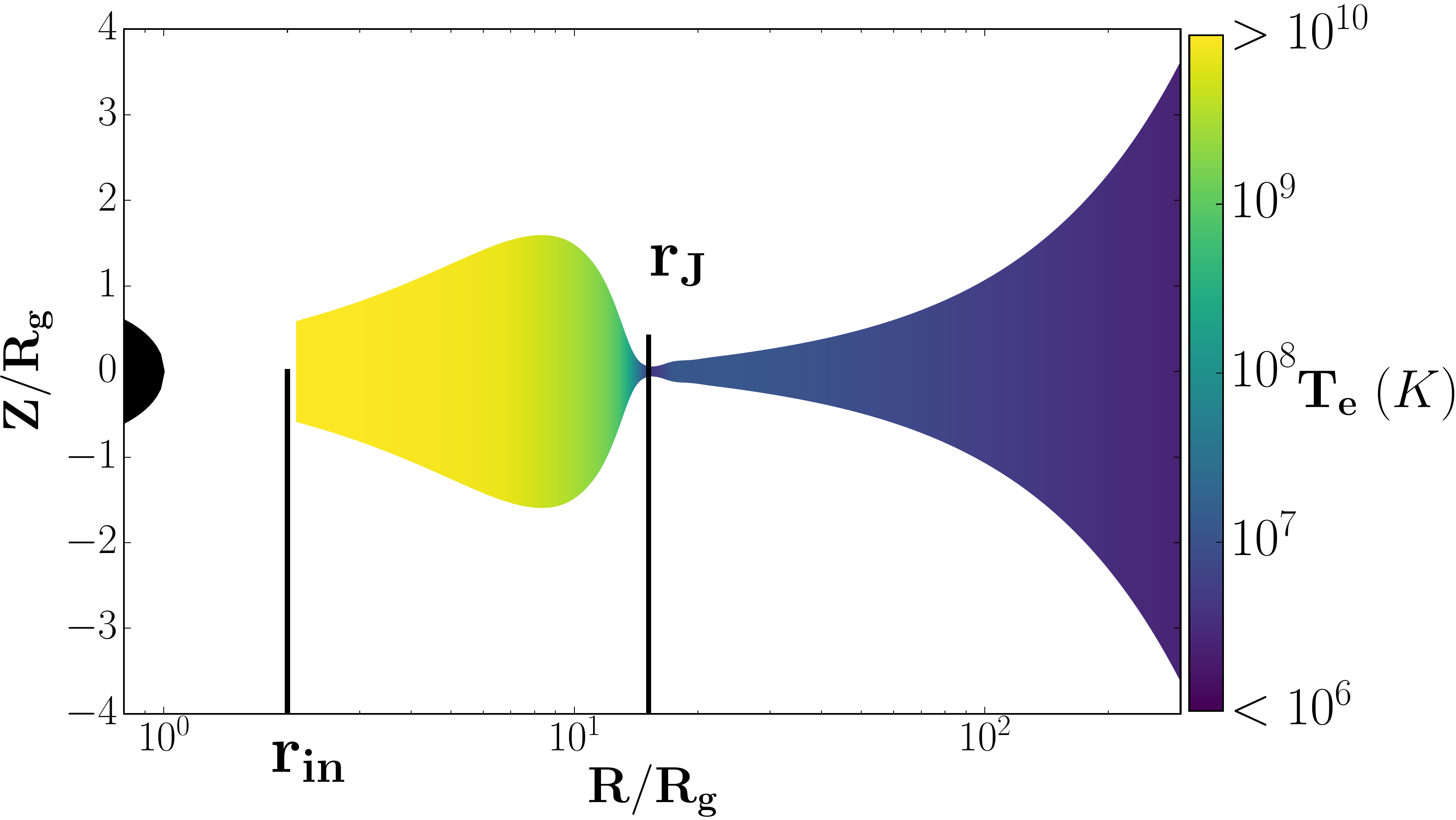}  
      \caption{Example of hybrid disk configuration in the JED-SAD paradigm. The inner disk regions are in a jet-emitting disk (JED) mode, up to a transition radius $r_J$, beyond which a standard accretion disk (SAD) is settled. The disk scale height $H(R)$ is accurately displayed, while colors corresponds to the central electronic temperature $T_e$ in Kelvin. The disk switches from an outer optically thick, geometrically thin jet-less disk to an inner optically thin, geometrically thick disk launching self-confined jets (not shown here). This solution has been computed for a transition radius $r_J=15$ and a disk accretion rate $\dot{m}_{in}=0.1$ at the disk inner radius $r_{in}=2$ (see section~\ref{sec:GeneralPpts} for more details). Other similar examples are shown in Fig.~\ref{fig:JoliesImages} for different pairs $(\dot{m}_{in}, r_J)$.}
         \label{fig:config}
\end{figure}

The hybrid disk configuration is composed of a black hole of mass $M$, an inner jet-emitting disk from the last stable orbit $R_{in}$ to the transition radius $R_{J}$ and an outer standard accretion disk from $R_{J}$ to $R_{out}$. The system is assumed to be at a distance $D$ from the observer. In the following, we adopt the dimensionless scalings: $r = R/R_g$, $h = H/R_g= \varepsilon r$, where $R_g = GM/c^2$ is the gravitational radius, $m = M/M_{\odot}$, and $\dot{m} = \dot{M}/\dot{M}_{Edd}$, where $\dot{M}_{Edd} = L_{Edd} / c^2$ is the Eddington accretion rate\footnote{Note that this definition does not include the accretion efficiency, usually of the order $\sim 10\%$ for a Schwarzschild black hole. This means that reaching Eddington luminosities would require $\dot{m}_{in} \gtrsim 10$ (see Fig.~\ref{fig:Lxmdot}).} and $L_{Edd}$ is the Eddington luminosity. Since GX 339-4 appears to be an archetypal object, we decided to concentrate only on this object. We thus use a black hole mass $m = 5.8$, a spin $a = 0.93$ corresponding to $r_{in} = 2.1$ and a distance $D = 8$~kpc \citep{Miller04}\footnote{See also \citet{Munoz08}, \citet{Parker16} or \citet{Heida17} for more recent estimations.}. All luminosities and powers are expressed in terms of the Eddington luminosity $L_{Edd}$. An example of disk configuration is shown in Fig.~\ref{fig:config} for $\dot{m}_{in} = \dot{M}_{in}/\dot{M}_{Edd} = 0.1$ and $r_J = R_J / R_g = 15$.\\

Our goal is then to compute, as accurately as possible, the radial disk thermal equilibrium from $r_{out}$ to $r_{in}$ by taking into account the known dynamical properties associated with each accretion mode. The inflow-outflow sctructure is described by the following midplane quantities:
\begin{eqnarray}
\mu &= &\frac{B_z^2/\mu_0}{P_{tot}} = \frac{B_z^2 / \mu_0}{P_{gas} + P_{rad}}   \nonumber \\
\xi &=& \frac{d \ln \dot{m}}{d \ln r} \nonumber \\
m_s &=& \frac{-u_R}{c_s} = \frac{-u_R}{\Omega_K H} =m_{s,turb} + m_{s, jet} = \alpha_{\nu} \varepsilon + 2 q \mu \\
 b &= & \frac{2 P_{jet}}{P_{acc, JED}} \nonumber
\end{eqnarray}
where $\mu$ is the disk magnetization, $\xi$ the local ejection index, $m_s$ the sonic accretion Mach number and $b$ the fraction of the JED accretion power $\displaystyle P_{acc, JED} = \left [ \frac{G M \dot{M}}{2 R} \right ]^{R_{in}}_{R_{J}}$ that is carried away by the jets in the JED. The parameter $q \simeq -B_\phi^+/B_z $ is the magnetic shear of the magnetic configuration and $B_\phi^+$ is the toroidal magnetic field at the disk surface \citep[see][for more details]{Ferreira97}. From the above radial distributions, we can deduce the expressions of the vertical magnetic field $B_z$, the accretion speed $u_R$ and the disk surface density $\Sigma$ as a function of $\dot{m}_{in}$ and the disk aspect ratio $\varepsilon=H/R$. The latter is obtained by solving the coupled energy equations for the ions and electrons in order to compute both electronic and ion temperatures (see paper II for a full description of the method).

\subsubsection{Inner JED}

Jet-emitting disks solutions from a large radial disk extent are known to exist in a restricted region of the parameter space \citep[][Fig. 2]{Ferreira97}. For simplicity, we assume that any given configuration is stationary and that parameters are constant. An extensive study of the thermal structure and associated spectra of JED can be found in paper II, where it is shown that the following set of parameters best reproduces XrB hard states, from low to very high luminosities: 
\begin{itemize}
\item $\mu= 0.5$: the disk magnetization $\mu$ has very little influence on X-ray spectra, because the synchrotron emission does not contribute much to the equilibrium (paper II). This value has thus been chosen to lie between the two extreme values allowed for JED solutions, namely $\mu_{min}=0.1$ and $\mu_{max}=0.8$ \citep{Ferreira97}.
\item $\xi=0.01$: the smaller the ejection index, the less mass is being ejected and the larger the asymptotic jet velocity. A value $\xi=0.01$ is consistent with mildly relativistic speeds \citep[e.g., the case study for Cyg X-1 in][]{Petrucci10}.
\item $m_s=1.5$: within the JED accretion mode, the jets torque is dominant and imposes $m_s= m_{s, jet} =2 q \mu $. The precise value of $m_{s,jet}$ depends on the trans-Alfv\'enic constraint, but accretion in a JED is always at least sonic and usually supersonic $m_{s,jet}>1$ \citep{Ferreira97}. In paper II we showed that a supersonic accretion with $m_s=1.5$ allows to reproduce luminous hard states. 
\item $b=0.3$: the fraction of the released accretion energy $P_{acc}$ transferred to the jets has been computed within self-similar models and goes from almost $1$ to roughly $0.2$ \citep{Ferreira97, Petrucci10}. The chosen value also appears as a good compromise and facilitates the reproduction of luminous hard states (paper II).
\end{itemize}

\noindent The fact that the disk accretion rate necessarily varies with the radius has been first introduced in accretion-ejection models by \citet{Ferreira93a}, in the context of magnetically driven jets, and later by \citet{BB99} in the context of thermally-driven outflows. In both cases, the disk ejection efficiency is characterized by the radial exponent $\xi$ in $\dot m (r) \propto r^\xi$. While it has been shown that magnetically-driven jets require $\xi < 0.1$ \citep{Ferreira97}, the values measured in many simulations is usually higher, lying between $0.5$ and $1$ \citep[][and references therein]{CasseKeppens04, YBW12, Yuan15}. It is somewhat troublesome that different simulations lead to a comparable value regardless of the strength of the magnetic field. Moreover, they were mostly done in the context of non-radiating hot accretion flows. But on the other hand, \citet{ZhuStone18} obtained $\xi \sim 0.003$ with an isothermal equation of state. Our guess is that this issue is not settled yet, especially given the extreme sensitivity of the disk ejection efficiency to the local thermodynamics \citep{Casse00b}. As discussed above, we thus simply assume a small value for $\xi$ that is compatible with the existence of relativistic jets.

\subsubsection{Outer SAD} \label{sec:innerSAD}

As argued in introduction, the outer disk regions are assumed to accrete under the SAD mode. This implies that the relevant torque is turbulent, probably due to the magneto-rotational instability (MRI hereafter). In this case, $m_{s,turb} = \alpha_{\nu} \varepsilon $. The Shakura-Sunyaev viscosity parameter $\alpha_{\nu}$ needs to be specified and we use $\alpha_{\nu}=0.1$ throughout this paper \citep{Hawley02, King04, Penna13}. The magnetization must be small enough to allow for the development of the MRI, we choose $\mu=10^{-3}$. As long as the SAD remains optically thick the value of $\mu$ does not affect our calculations of the SAD thermal equilibria. \\

By definition, no jets are present in a SAD. This translates into $\xi_{SAD}=0$ (no mass loss) and $b_{SAD}=0$ so that all released energy is either radiated or advected by the flow. Doing so, we neglect the potential presence of winds usually observed in XrB, especially at high luminosities \citep{Ponti12, Tetarenko16}. This assumption sounds reasonable for two reasons. First, although mass loss from turbulent disks is indeed possible and actually observed in MHD simulations \citep[e.g.,][]{Proga00, BaiStone13, Suzuki14, Bethune17, ZhuStone18}, these magneto-thermally driven flows carry away a negligible fraction of the disk angular momentum and released accretion energy, introducing thereby no significant change in the disk structure. Second, the mass rate feeding the inner JED is simply $\dot{m}(r_J)= \dot{m}_{in} (r_J/r_{in})^{\xi}$ and is independent from $\xi_{SAD}$. Increasing $\xi_{SAD}$ up to say, $0.5$, would imply a strong increase of the disk accretion rate in the regions beyong $r_J$ up to $r_{out}$. This would of course lead to a significant change of the emitted spectrum from these outer regions, but with no detectable counterpart \citep[see however][and references therein]{Susmita16} as long as the disk remains optically thick, which is the case here.


\subsubsection{JED-SAD radial transition}
\label{sec:inter}

We examine here some properties of the transition, assuming that it occurs over a radial extent of the order of a few local disk scale heights. \\

The first striking property is the existence of a trans-sonic critical point near $R_J$. Indeed, while the accretion flow is subsonic in the SAD with a Mach number $m_{s,SAD} = m_{s,turb} = \alpha_{\nu} \varepsilon_{SAD} \ll 1$, it is supersonic in the inner JED with $m_{s,JED} = m_s = 1.5$. This property is a natural consequence of the transition from a turbulent "viscous" torque acting within the outer SAD to a dominant jet torque in the JED. Since the disk is assumed to be in a steady-state, the continuity of the mass flux $\dot{M}_{JED}=\dot{M}_{SAD}$ must be fulfilled at the transition radius. Given the difference between the JED and SAD sonic Mach number, this implies a drastic density decrease $\displaystyle \Sigma_{JED}/\Sigma_{SAD} \sim \alpha_{\nu} \varepsilon_{SAD}^2/\varepsilon_{JED} \ll 1$ between the SAD and the JED. The Thomson optical depth being defined by $\tau_T \propto \Sigma$, this density drop therefore implies a huge drop in the disk optical depth. Thus, a dynamical JED-SAD radial transition naturally goes with an optically thin - optically thick transition (see paper II). 
 
The second striking property is the possible existence of a thin super-Keplerian layer between the JED and the SAD. In the outer SAD, the disk is slightly sub-Keplerian with a deviation due to the radial pressure gradient and of the order of $\varepsilon_{SAD}^2$. Within the JED, the much larger magnetic radial tension leads to a larger deviation of the order of $\mu \varepsilon$ \citep{Ferreira95}. This requires that the radial profile $\Omega(R)$ has two extrema (with $d\Omega/dR \simeq 0$). Since all the disk angular momentum is carried away vertically in JEDs, there is no outward angular momentum flux into the SAD at $R_J$. This translates into a "no-torque" condition for the SAD. Such a situation has already been discussed in the context of a radial transition between an outer cold (SAD) and inner hot (ADAF) accretion flows, leading to a super-Keplerian layer \citep{Honma96, Abramowicz98}. It is, however, not clear whether such a thin layer would still be present in our context given the existence of magnetic forces. But these two extrema of the angular velocity clearly define the radial end points of each dynamical (SAD or JED) solution, and the trans-sonic transition occurring in-between. \\

Constructing a dynamical solution describing the radial transition between a SAD and a JED is beyond the scope of the present paper and will be studied elsewhere.
From now on, we assume that the two accretion modes can always be matched at a transition radius $R_J$. The calculation of the global disk equilibrium can then be undergone using $\dot{m}_{in} = \dot{M}_{in} / \dot{M}_{Edd}$ and $r_J = R_J / R_g$ as independent variables.



\subsection{Thermal structure of hybrid JED-SAD configurations}
\label{sec:discussionJED}

As described in paper II, our accretion flow is locally described by a two-temperatures ($T_e$, $T_i$), fully ionized plasma of densities $n_e = n_i$, embedded in a magnetic field $B_z$. We recall here the main equations used to compute the thermal equilibrium (see paper II for full explanations). The electron and proton temperatures are computed at each radius using the coupled steady-state local energy balance equations
  \begin{eqnarray}
     \left( 1-\delta \right) \cdot q_{turb} &=&  q_{adv,i} + q_{ie} ~~~~~~~~~~~~~~~~~~~~ \text{ions} \label{eq:IONS} \\
      \delta \cdot q_{turb} &=& q_{adv,e} - q_{ie} + q_{rad} ~~~~~~~~~ \text{electrons} \label{eq:ELECTRONS}
   \end{eqnarray}
where the local heating term, of turbulent origin, varies according to the radial zone considered. Within the JED in $R<R_J$, it gives
\begin{equation}
q_{turb,JED} =   (1-b) (1-\xi) \frac{G M \dot{M}(R)}{8 \pi H R^3} 
\end{equation}
whereas within the SAD in $R>R_J$ it is 
\begin{equation}
q_{turb,SAD}  =   \frac{3 G M \dot{M}(R)}{8 \pi H R^3} \left( 1 - \sqrt[]{R_{J}/R} \right) \label{eq:qturbSAD}
\end{equation}
making use of the no-torque condition imposed at $R_{J}$ and a constant $\dot{M}$ at the transition (section~\ref{sec:inter}). In principle, the released turbulent energy could be unevenly shared between ions and electrons by a factor $\delta$. Throughout this paper, we use $\delta=0.5$ \citep[see paper II,][section 2.3 and references therein]{Yuan14}. The other terms appearing in Eq.~(\ref{eq:IONS}) and (\ref{eq:ELECTRONS}) are the ion (electron) advection of internal energy $q_{adv,i}$ ($q_{adv,e}$), the Coulomb collisional interaction between ions and electrons $q_{ie}$ and the radiative cooling term due to the electrons $q_{rad}$. This term, as well as the radiation pressure $P_{rad}$ term, is computed using a bridge function allowing to accurately deal with both the optically thin and the optically thick regimes (paper II). The optically thin cooling regime $q_{thin}$ is computed with the BELM code \citep{Belmont08}, which includes Compton scattering, emission and absorption through bremsstrahlung and synchrotron processes.

The thermal equilibrium of the SAD region is well known. For large accretion rates, required for outbursting XrBs, the disk is mostly in the optically thick, geometrically thin cold regime ($T_e \sim 10^5-10^7$~K). 
Heating of the SAD surface layers by the hard X-rays emitted by the inner JED might produce some disk evaporation \citep[][]{Meyer94, Meyer00, Liu05, Meyer05}. This would require to solve the 2D (vertical and radial) stratification of the disk, which is beyond our vertical one-zone approach. However, this is not expected to be crucial for the scenario depicted here. We therefore neglect the feedback of the inner JED over the outer SAD structure. In this approximation and since the resolution is computed outside-in, the temperature of the outer SAD does not depend on any assumption made on the inner JED.\\

On the contrary, the effects of the outer SAD on the inner JED are twofold and cannot be neglected. The first effect is the cooling due to advection of the outer cold material into the JED. Indeed the local advection term $q_{adv}$ can be either a cooling or a heating term, depending on the sign of the radial derivatives. This is self-consistently taken into account in our code (see Eq. (12) and Appendix A.2 in paper II).
The second effect results from the Compton scattering of the SAD photons on the JED electrons. This effect occurs whenever the JED is in the optically thin, geometrically thick thermal solution. This effect was not taken into account in paper II. \\

Illumination is estimated from the SAD properties, using the following geometrical prescriptions. Outside of the transition radius $R_J$, the disk luminosity is $\displaystyle L_{SAD} = G M \dot{M} / 2 R_J \simeq 4 \pi R_*^2 \sigma T_s^4$, where $T_s= T_{eff}(R_*)$ is the effective temperature at the radius $R_*$ where $T_{eff}$ reaches its maximum \citep{Frank02}. JED solutions being optically thin (or slim at worse), we assume that the radiation field in the region below $R_J$ is well described by the average energy density
\begin{equation} 
U_{rad} = \omega \frac{L_{SAD}}{4\pi R_J^2 c} \label{eq:Urad}
\end{equation}
where $0 < \omega < 1$ is a geometrical dilution factor that describes the fraction of the SAD power that irradiates the region below $R_J$ (see section~\ref{sec:EffectOfOmega}). This applies to the bolometric luminosity, but also to the luminosity in any given energy band.

This prescription allows to compute all properties of the illumination field. This radiation is then provided to the BELM code as an external source of seed photons, and the associated cooling and reprocessed spectrum are computed. More precisely, the JED is divided in many spheres of radius $H$ in which radiation processes are computed (see paper II for more details). Here, each sphere of radius $H$ receives the power
\begin{equation} 
L_{s} = U_{rad} 4\pi H^2 c = \omega (H/R_J)^2L_{SAD} \label{eq:Ls}
\end{equation}
\noindent where the value of $\omega$ is discussed in section~\ref{sec:EffectOfOmega}.

\begin{figure*}[t]
  \centering
  \includegraphics[width=\textwidth]{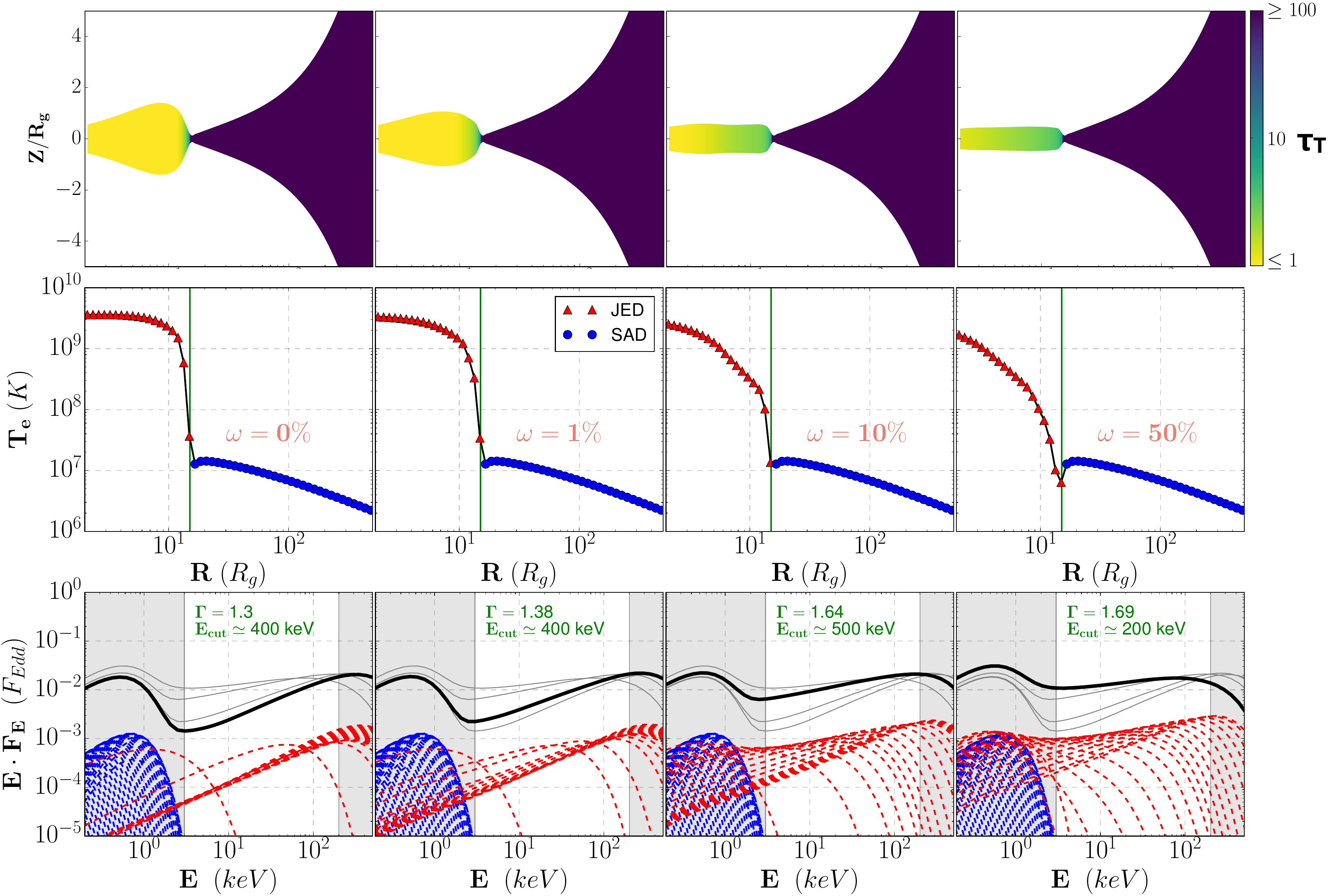}
   \caption{Effect of the external Comptonisation of the SAD photon field on the inner JED for $\dot{m}_{in} = \dot{M}_{in} / \dot{M}_{Edd} = 1$ and $r_J = 15$ (green vertical line on the top figures), with different dilution factors, from left to right $\omega = 0,~1,~10,~50 \%$. Top panels show the disk aspect as well as its Thomson optical depth in colors. Middle panels show the electron temperature as function of radius. Bottom panel displays the local spectra emitted by each radius (dashed lines) and the corresponding total disk spectrum (black solid line). For comparison, in each panel we have overplotted in gray solid lines the total spectra obtained in the other three panels. The spectra are given in Eddington fluxes $F_{Edd} = L_{Edd} / 4 \pi D^2$ units for GX 339-4 (see section~\ref{sec:GeneralPpts}). The blue lines and dots correspond to the SAD zone, while the red lines and triangles represent the JED. The white part of the spectra shows the $3-200$~keV energy range. Approximate values of the photon index $\Gamma$ and energy cutoff $E_{cut}$, derived by comparison with a simple cutoff power-law model in this energy range, are indicated on each plot.}
     \label{fig:EffectOfOmega}
\end{figure*}

\subsection{Effect of an external illumination on the JED thermal structure} \label{sec:EffectOfOmega}

Adding an outer standard accretion disk may have a colossal impact on the inner hot JED, depending on the transition radius $r_J = R_J / R_g$. For large values of $r_J$ (say larger than $50-100$), the power emitted by the SAD is too low to affect the global disk spectrum. But this is no longer the case when the transition radius becomes smaller. 
Besides, the geometrical dilution factor $\omega$ used in Eq.~(\ref{eq:Urad}) plays an important role. It is however quite tricky to obtain an accurate estimate of its value within our framework. It depends on the solid angle under which the SAD photosphere is seen by the JED and corresponds thereby to the fraction of the SAD photons that are intercepted by the JED. 

Considering a spherical hot corona of radius $r_J$, centered on the black hole and embedded in an infinitely thin disk, former studies \citep[e.g.,][]{Zdz99, Ibragimov05} led to $\omega \sim 2 - 25 \%$ depending on the dynamical ("no-torque", "torque") hypothesis made at $r_J$. The inner JED accretion flow is clearly different from a sphere of radius $r_J$ (see Fig.~\ref{fig:config}), which might naively suggest a value smaller than the above estimates. Moreover, Compton cooling also should be a function of the radius within the JED, and in the case of the geometry shown in Fig.~\ref{fig:config}, we could expect $\omega$ to decrease with decreasing radius. \\

There are however numerous effects that should magnify $\omega$. 
First, the SAD is clearly flared, thus the infinitely thin disk approximation generally used is rather crude and tends to decrease the value of $\omega$ \citep[see, e.g.,][]{Meyer05, Mayer07}.
Second, although not considered in the litterature, the photons emitted radially by the innermost region of the SAD are also expected to radiate directly towards the JED (see Fig.~\ref{fig:EffectOfOmega} top panel). Indeed, the photosphere $\tau_T = 1$ necessarily crosses the disk midplane near $r_J$, allowing cold SAD photons to enter directly into the outer parts of the JED (and not only from the SAD surfaces). This should be responsible for another radiative contribution. The estimation of the corresponding photon flux emitted by the SAD and entering the JED in such a way is quite complex. It would require to solve the full (radial + vertical) radiative transfer problem to determine the photosphere properties of the SAD close to $r_J$. While this is far beyond the goal of the present paper, this effect should be similar to estimations\footnote{Again, in the case of an infinitely thin disk, and for a disk penetrating the hot corona in $r_J$ \citep[see again][]{Zdz99}, resulting in $\omega$ of the order of tens of percent depending on how far the disk penetrates inside the corona.}. 
Third, the reprocessing of the X-ray emission from the JED inside the disk will also naturally increase the SAD emission \citep{Poutanen17}. In our model, we only take into account the intrinsic disk emission, but X-ray reprocessing can be mimicked by increasing $\omega$.
Fourth, another important effect that should be taken into account is the gravitational light bending close to the black hole. This should strongly magnify the flux of disk photons impinging the JED in comparison to the Newtonian situation where they would mainly escape away from the disk. This effect should depend on the transition radius $r_J$ as well as the radial position inside the JED. Ray tracing simulations are required here for a rigorous computation, and here again this is out of the scope of the present paper. But this effect could be the dominant one especially for small radii inside the JED or for small $r_J$, since the closer to the black hole the stronger the light bending effect. This could result in a factor of a few to be applied to the number of SAD photons entering the JED in comparison to the absence of light bending \citep[see, e.g.,][]{Miniutti03}. All included, we believe that $\omega$ of the order of a few tens of percent seems rather reasonable. \\

We report in Fig.~\ref{fig:EffectOfOmega} the JED radial temperature distribution, as well as the corresponding spectral energy distribution (SED), for a constant $\omega$ varying from $0$ to $50 \%$, the two physical extremes values for this parameter. Increasing $\omega$ obviously decreases the JED temperature and softens the SED. The variation is quite important between $\omega=0\%$ and $50 \%$. Using a simple power-law model we find a spectral softening of $\Delta \Gamma \simeq 0.4$ of the resulting power-law, along with a modification of its energy cutoff, from $E_{cut} \simeq 500$~keV to $E_{cut} \simeq 200$~keV.

Clearly, this effect can not be neglected, as the value of $\omega$ has an important impact on the spectra. In this article, unless otherwise specified, we use $\omega = 0.2$. This value appears to be close to the upper limit for previous estimations, but, considering the number of assumptions diminishing this value, we thought this was a good compromise. We note however that this does not mean that the inner JED captures $20\%$ of the SAD luminosity, as there is still a factor $\left( H/R_J \right)^2$ in Eq.~(\ref{eq:Ls}).

\section{Synthetic observations: X-ray disk spectra and radio jet emission} \label{sec3}
\label{sec:Syntheticspectra}

\subsection{From theoretical SEDs to simulated data}
\label{sec:Py2Xspec}

This work aims at providing synthetic hardness-intensity diagrams, or more precisely, disk fraction luminosity diagrams \citep[DFLD, see, e.g.,][]{Kording06, Dunn10}. To that purpose, our synthetic spectra must be processed in a way similar to observational data to derive the disk and power-law components from the fits, and place the corresponding points in a DFLD. This procedure, too rarely performed, is mandatory as we intend to compare our synthetic data to observations.

\begin{figure}[h!]
  \includegraphics[width=1\linewidth]{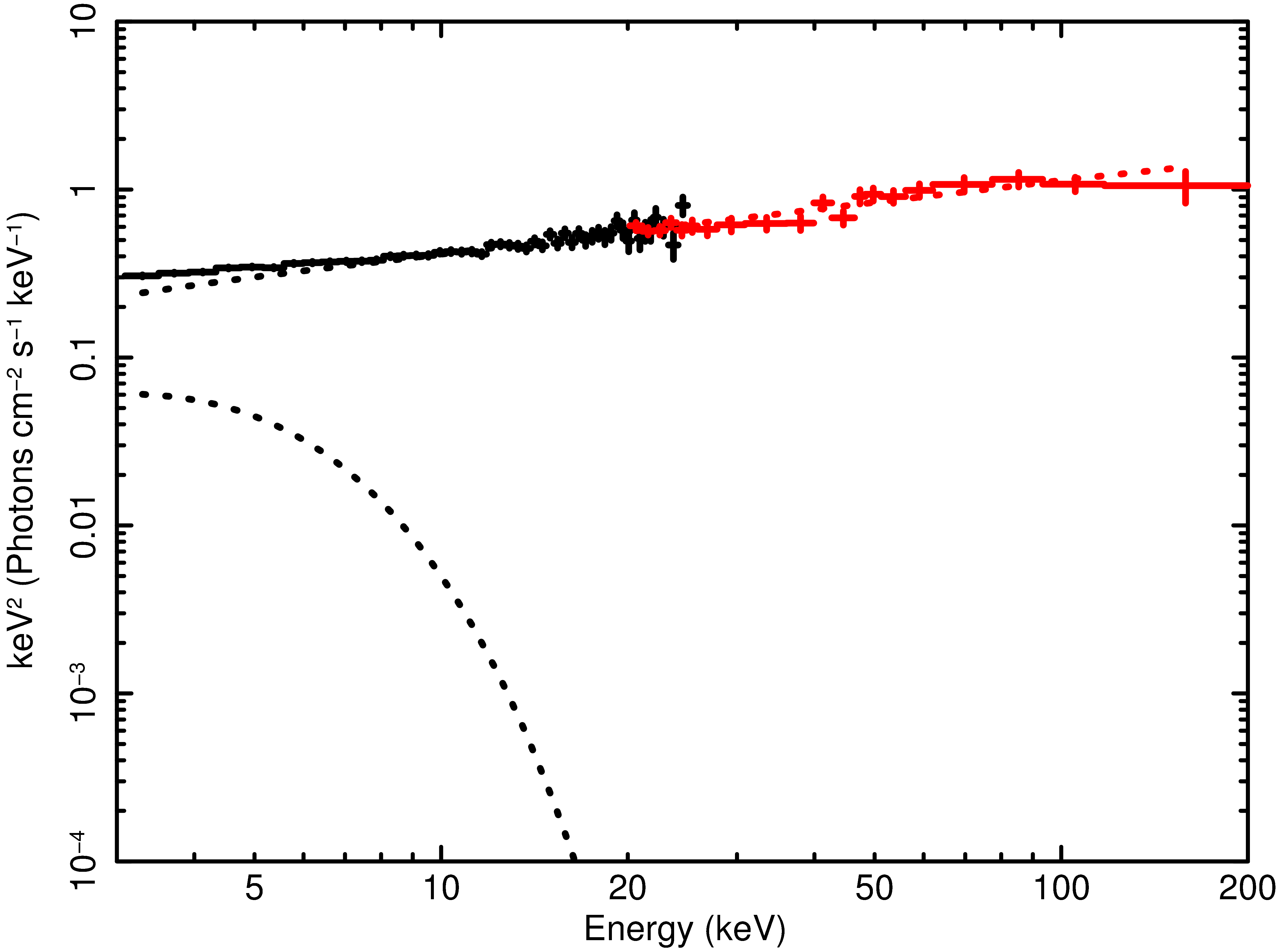} 
   \caption{Example of simulated \textsc{RXTE/PCA} ($3-25$~keV, in black) and \textsc{RXTE/Hexte} ($20-200$~keV, in red) data sets from the theoretical SED produced by the JED-SAD configuration shown in Fig.~\ref{fig:EffectOfOmega}, third panel. The dotted lines are the power law and disk components corresponding to the best fit model. See section~\ref{sec:Py2Xspec} for more details.}
     \label{fig:Xspec}
\end{figure}

From our theoretical SED, an \textsc{XSPEC} table model was first built using the \textsc{flx2tab}\footnote{https://heasarc.gsfc.nasa.gov/ftools/caldb/help/flx2tab.html} command of \textsc{ftools}\footnote{https://heasarc.gsfc.nasa.gov/ftools/}. Disk inclination was ignored for simplification but we add background and galactic absorption (\textsc{wabs} model in \textsc{XSPEC} with $N_H = 4 \times 10^{21}$~cm$^{-2}$, see \citealt{Dickey90, Dunn10, Clavel16}). Then we simulated RXTE/PCA and HEXTE spectrum with the \textsc{XSPEC} simulation command \textsc{fakeit}. We use exposure times $t_{exp}$ between $1$ and $10$~ks depending on the model flux in the $3-200$~keV band in order to have a reasonably good signal-to-noise ratio. In this article, we use $t_{exp} = 10$~ks for the quiescent state (section~\ref{sec:XrB5states}) and $t_{exp} = 1$~ks elsewhere. In Fig. \ref{fig:Xspec}, we plot for example the simulated spectrum from the theoretical SED produced by the JED-SAD configuration with $\omega = 0.1$ (Fig.~\ref{fig:EffectOfOmega}, 3rd panel). \\

Each simulated spectrum is then fitted with three different models (a): \textsc{wabs} $\times$ (\textsc{cutoffpl} + \textsc{ezdiskbb}), (b): \textsc{wabs} $\times$ \textsc{cutoffpl} or (c): \textsc{wabs} $\times$ \textsc{ezdiskbb}. We keep the best one according to a \textsc{ftest} procedure \cite[see, e.g.,][section 3.2]{Clavel16}. In the example shown in Fig.~\ref{fig:Xspec}, model (a) gives the best fit with a reduced $\chi^2_{red} = 248.4 / 213 = 1.17$. The best fit parameters are a disk blackbody temperature $T_{in} = 1.0 \pm 0.3$~keV, a photon index $\Gamma = 1.59 ^{+0.03}_{-0.06}$ and a lower limit in cutoff $E_{cut} > 400$~keV. These values are consistent with simple estimations performed on the theoretical data $\Gamma = 1.64$ and $E_{cut} \simeq 500$~keV. The corresponding best fit disk and power-law components are plotted in dashed lines in Fig.~\ref{fig:Xspec}.

The total unabsorbed disk luminosity $L_{Disk}$ and unabsorbed power law luminosity $L_{PL}$ are computed in the $3-200$~keV range. In the example shown in Fig.~\ref{fig:Xspec}, the powerlaw fraction, defined by $PL_f = L_{PL} / (L_{PL} + L_{Disk})$ is equal to $0.99 \pm 0.01$ and the total flux in the $3-200$~keV band is $L_{tot} = L_{PL} + L_{Disk} = 2.9 \pm 0.1 ~ \% L_{Edd}$.

\subsection{Hard tail}
\label{sec:HardTail}

\begin{figure}[h!]
  \includegraphics[width=1\linewidth]{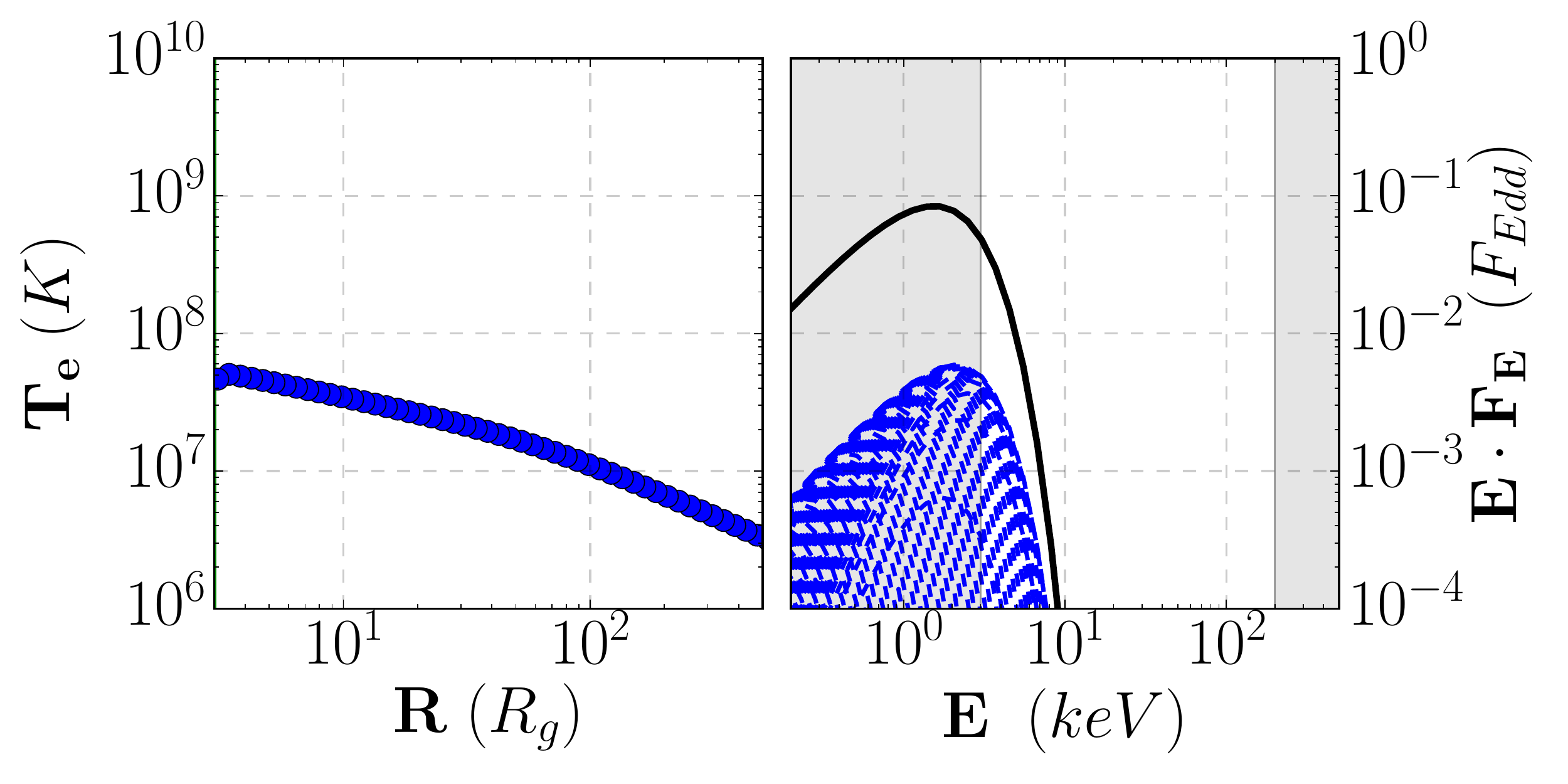}
  \includegraphics[width=1\linewidth, trim = 0.5cm 1.3cm 2.5cm 2.5cm, clip]{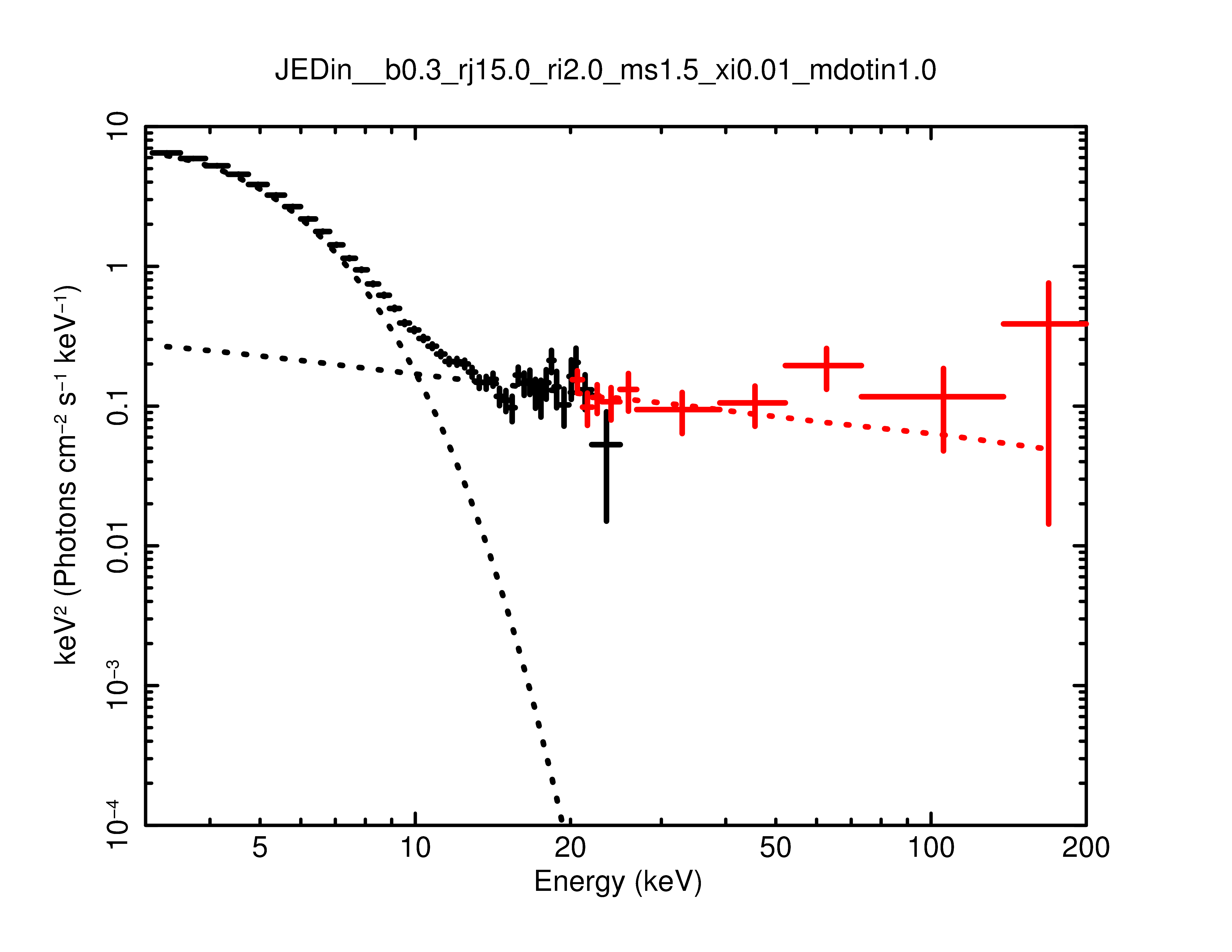}
   \caption{Electron temperature (top-left) and theoretical spectrum (top-right) of the configuration $\dot{m}_{in} = 1$, $\mu_{SAD} \ll 1$, $\alpha_{\nu} = 0.1$, $r_J = r_{in}$. Each annulus is displayed as a blue dot, its associated spectrum in blue dashed lines and the total disk spectrum in black solid line. In the bottom panel, final faked and fitted data after the addition of the hard power-law tail, black for \textsc{PCA} and red for \textsc{\textit{HEXTE}}. Dashed lines show the best fit obtained with \textsc{XSPEC}, see section~\ref{sec:HardTail} for details.}
     \label{fig:HardTail}
\end{figure}

It is well know that soft states show non thermal tails generally observed above few keV \citep{McConnell02,Remillard06}. Although uncertain, these tails are thought to be produced by a population of non-thermal electrons \citep[see][for few investigations]{Galeev79, Gierlinski99} that are not taken into account in our model. In order to reproduce such soft states, we added a power-law component to the synthetic spectra each time the fitting procedure favors a pure blackbody emission (model (c)) and the new data were re-fitted with a modified model (c) \textsc{wabs} $\times$ (\textsc{pl} + \textsc{ezdiskbb}) model. The photon index of this power-law component is set to $\Gamma=2.5$ and it is normalized in order to contribute to a fixed fraction of the $3-20$~keV energy range \citep[typically between $1\%$ and $10\%$, ][]{Remillard06}. \\

An example is shown in Fig.~\ref{fig:HardTail}. The theoretical model plotted on top of this figure corresponds to a SAD extending down to $r_J = r_{in}$ (no JED), with $\dot{m}_{in} = 1$. A fit with an absorbed disk component (model (c)) gives a disk temperature $T_{in} = 0.97 \pm 0.01$~keV and a total flux $L_{tot} = 4.1 \pm 0.1 \% L_{Edd}$ in the $3-200$~keV band. The data simulated in XSPEC include the additional power-law tail (Fig.~\ref{fig:HardTail}, bottom panel). The best fit with new model (c) then gives $T_{in} = 0.97 \pm 0.01$~keV, $\Gamma = 2.5 \pm 0.2$ and $L_{tot} = 4.5 \pm 0.1 \% L_{Edd}$ with a reduced $\chi^2_{red} = 214.5 / 213 = 1.01$. We note that this procedure slightly increases the total flux detected in X-rays due to the addition of the hard power-law tail.

\begin{figure*}[h!]
  \centering
  \includegraphics[width=1.\textwidth]{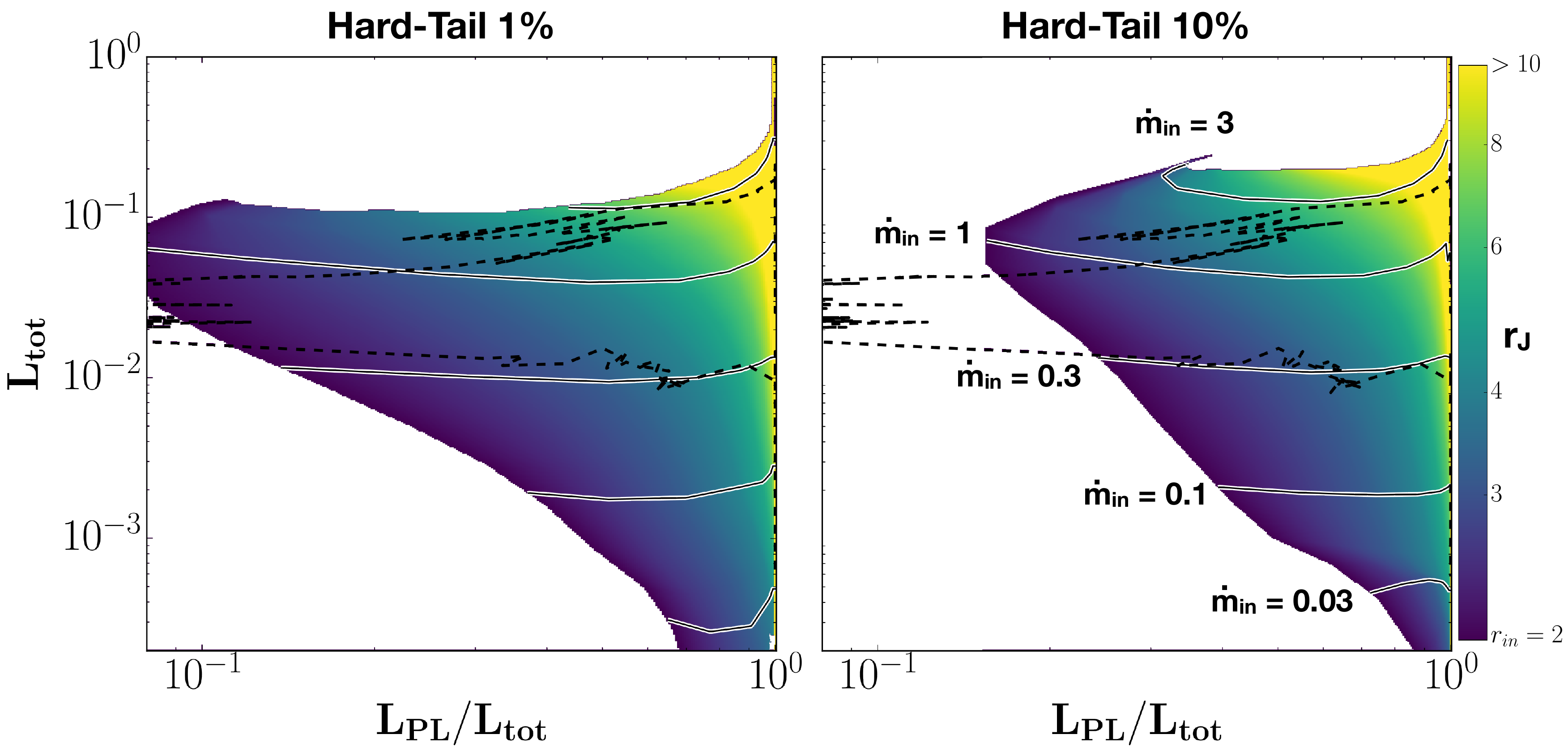}  
   \caption{Total, disk + power-law, luminosity $L_{tot}=L_{disk} + L_{PL}$ in the $3-200$~keV energy range (in Eddington luminosity unit) is shown in function of the power-law fraction $L_{PL}/L_{tot}$. Each point within this plot corresponds to a fully computed and then XSPEC processed hybrid JED-SAD configuration. Contours (black solid lines) are for a constant disk accretion rate $\dot{m}_{in}$ while the color background display the disk transition radius $r_J$. Dashed black line shows the 2010-2011 cycle of GX 339-4. XSPEC fits were done with a hard tail level of $1\%$ (left) and $10\%$ (right). See section \ref{sec:EffectOfRjMdot} for a description of the figure.}
     \label{fig:EffectOfRj}
\end{figure*}

\subsection{Jet power and radio luminosity}
\label{sec:Pjet}

In a JED, the jets power available is a given fraction of the accretion power
\begin{eqnarray}
  P_{jets} &=& b  P_{acc, JED} = b  \left [ \frac{G M \dot{M}}{2 R} \right ]^{R_{in}}_{R_{out}} \nonumber \\
  		       &=& \frac{b}{2} \frac{\dot{m}_{in}}{r_{in}} \left( 1 - \left( \frac{r_J}{r_{in}} \right)^{\xi-1} \right) L_{Edd} \label{eq:Pjet}
\end{eqnarray}
Assuming that $b$ is roughly a constant throughout an entire evolution, the jets power depends on both $\dot{m}_{in}$ and $r_J$. We follow the computations of \citet{Heinz03} to deduce the expected radio luminosity emitted by one jet component. It assumes that the jet emission is explained by self-absorbed synchrotron emission of non-thermal particles along the jet. We need however to modify their equations in order to account for the finite radial extent of the JED, imposing a finite radial extent of each jet. This leads to the following expression (see Appendix~\ref{sec:JetRadio} for more details):
\begin{equation}
\frac{\nu_R L_{R}}{L_{Edd}} \simeq f_{R}\,   m^{\beta-1} \,  r_{in}^{- \frac{6p+49}{4p+16}} \, \dot{m}_{in}^\beta \,  r_J^{\frac{p+9}{p+4}}\,  \left (1- \frac{r_{in}}{r_J} \right )^{\frac{5}{p+4}}
\label{eq:Fr}
\end{equation}
where $L_R= 4\pi D^2 F_R$ is the monochromatic power emitted at the radio frequency $\nu_R$ from an object at a distance $D$. The parameter $f_R$ is a normalization constant, $\beta= \frac{2p+13}{2p+8}$ and $p$ is the usual exponent of the non-thermal particle energy distribution. In the standard case with $p=2$ and $r_J$ constant, one gets the \citet{Heinz03} dependencies, namely a radio power $\nu_R L_{R} \propto \dot{m}_{in}^{17/12}$.\\

Our model then provides naturally both $L_R$ and $L_X$. Indeed, for any given set of parameters $(\dot{m}_{in}, r_J)$, we can compute $L_X$ from our simulated SED, whereas an estimate of the radio luminosity can be obtained using Eq.~(\ref{eq:Fr}). This is discussed in section~\ref{sec4}.

\section{Reproducing typical XrB behavior: DFLD and canonical spectral states} \label{sec4}

In this section, we show that hybrid JED-SAD configurations can, in principle, reproduce the outbursting cycles of XrBs by varying only two parameters, the disk accretion rate $\dot{m}_{in}$ and the transition radius $r_J$. This is done in two steps. First, we need to find which ranges in $\dot{m}_{in}$ and $r_J$ allow to cover the full DFLD. However, this diagram includes only information on X-ray emission while a cycle also deals with jet production and quenching. We thus require that the same framework reproduces radio emission at the correct level. Then, as a second step, we define five canonical spectral states characteristic of an XrB spectral evolution during an outburst and show more precisely how well our framework is able to reproduce them.

\subsection{Disk fraction luminosity diagram}
\label{sec:EffectOfRjMdot}

We perform a large parameter survey for $\dot{m}_{in} \in [0.01, ~ 10]$ and $r_J \in [r_{in},~ \sim 50 r_{in}] = [2, ~ 100]$. We compute the whole thermal structure and corresponding theoretical SED for each pair $(\dot{m}_{in}, r_J)$, and we fit as described in section~\ref{sec:Syntheticspectra} to get the corresponding position in the DFLD. The fits\footnote{Only fits with $\chi^2_{red} < 3$ have been displayed here. Few fits $(\dot{m}_{in} \gtrsim 5,~ r_J = r_{in})$ require the addition of a second blackbody component to better describe their spectral shape, but their position (top-left) beyond the extension of usual DFLDs makes them meaningless in the current study.} are shown in Fig.~\ref{fig:EffectOfRj}. The smoothness of our DFLD is indicative of the absence of spectral degeneracy in our modeling (Fig.~\ref{fig:EffectOfRj}).
In this figure, the mean transition radius is color-coded and the accretion rate is shown in contours of constant values. For comparison, the 2010-2011 outburst of GX 339-4 is overplotted in black dashed line\footnote{The spectral analysis of the 2010-2011 outburst was done in the $3-25$~keV energy range (\textsc{RXTE/PCA}), but the models were integrated in the $3-200$~keV range and not the usual ranges \citep[e.g.,][]{Dunn10}\label{fnt:ranges}}.


Fig.~\ref{fig:EffectOfRj} shows that we can cover the whole domain usually followed by XrBs within our framework. Concerning the hard states, we are able to reproduce their evolution up to high luminosities. Concerning the soft states, their position in the DFLD depends on the amplitude of the additional hard tail. With a hard tail representing $10\%$ of the flux in the $3-20$~keV energy range throughout the cycle, we can only reproduce soft states with $L_{PL}/L_{tot} > 0.1$ (Fig.~\ref{fig:EffectOfRj}, right). Softer states, populating the very left part of the DFLD, require a hard tail flux fraction lower than $1\%$ (Fig.~\ref{fig:EffectOfRj}, left).

As expected, the accretion rate is mainly responsible for the global X-ray luminosity of the system, leading to almost horizontal isocontours for $\dot{m}_{in}$ in the DFLD. Indeed, the higher the accretion rate, the higher the total available accretion energy $P_{acc} \propto \dot{m}_{in}$. However, part of the energy can be advected, which explains the sharp variations of the isocontours at high luminosities (see discussion below). The effect of the transition radius $r_J$ follows the predictions of paper I. At large transition radii, most of the emission originates from the JED, as the outer SAD has no detectable influence. These solutions display power-law spectra for all accretion rates (see paper II). This is the reason why they appear on the right-side of the DFLD. At small transition radii, two effects appear in the RXTE energy range of our simulated data. First, as its temperature and flux increase with decreasing $r_J$, the SAD blackbody emission starts appearing in the SED around $3$~keV. Second, the closer the SAD, the stronger its illumination becomes, cooling down the inner JED and producing softer spectra. Combining these two effects leads to a disk dominated spectrum, with a power-law fraction becoming entirely dominated by the high energy tail when $r_J \rightarrow r_{in}$. \\

\subsection{$L_X$ dependencies on $\dot{m}_{in}$}

\begin{figure}[h]
  \centering
  \includegraphics[width=\columnwidth]{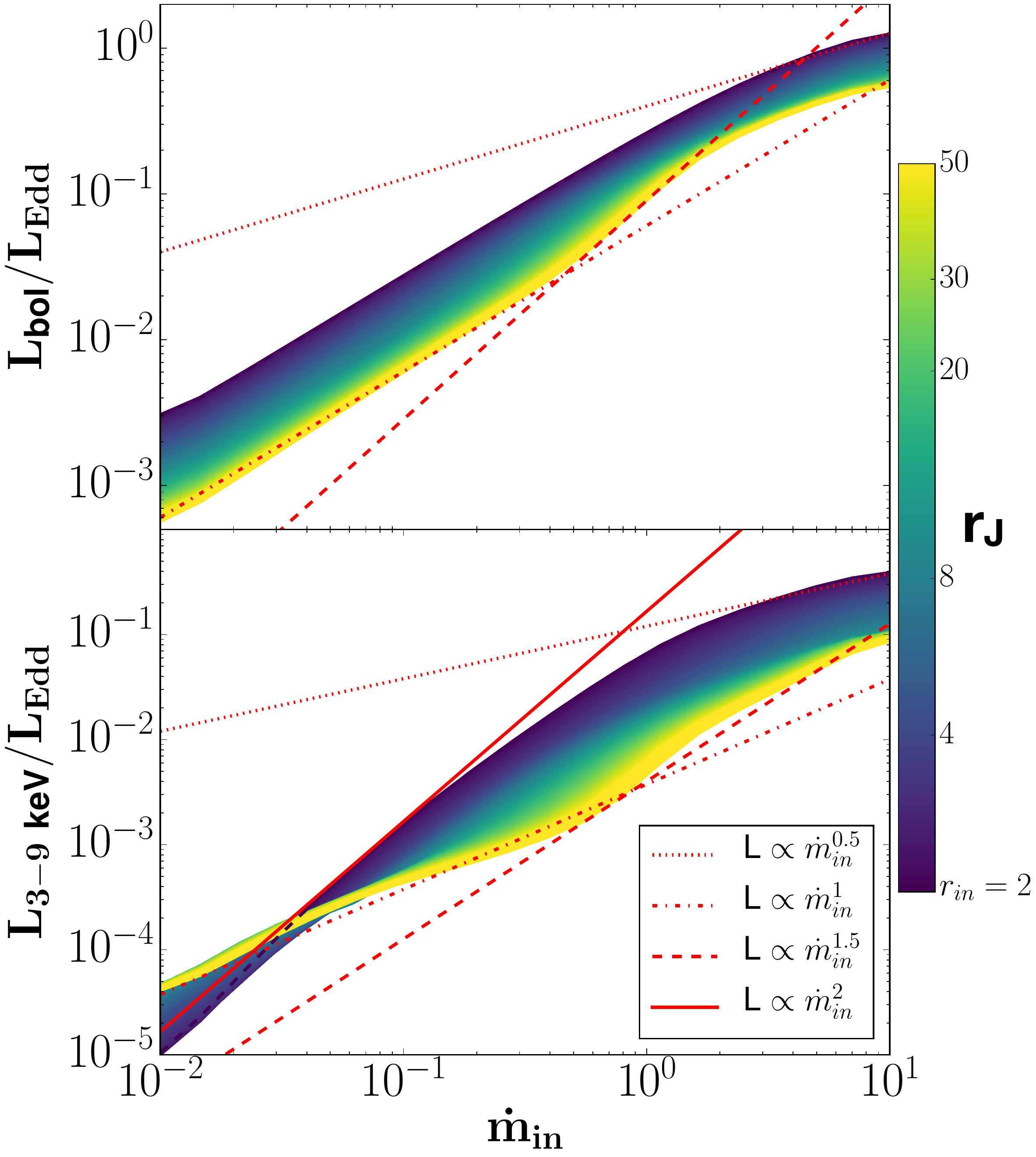}
   \caption{Bolometric (top) and $3-9$~keV (bottom) luminosities in function of the mass accretion rate $\dot{m}_{in}$ onto the black hole. This plot is extracted from Fig.~\ref{fig:EffectOfRj}, done with a $10\%$ hard tail (right). The colors are for different values of the transition radius $r_J$. Four different $L \propto \dot{m}_{in}^{\alpha}$ regimes are shown. Also, the $r_J = r_{in}$ has been drawn in dashed black to be visible at low accretion rate in the bottom panel.}
     \label{fig:Lxmdot}
\end{figure}

Figure~\ref{fig:Lxmdot} shows the bolometric ($L_{bol}$, top) and $3-9$~keV ($L_{3-9}$, bottom) luminosity deduced from our synthetic SED as a function of the accretion rate at the inner radius $\dot{m}_{in}$. The colors correspond to different transition radius $r_J$. 
This figure illustrates different concerns about the radiative efficiency of accretion flows. \\

On the top panel and in the SAD mode ($r_J = r_{in} = 2$), the bolometric luminosity follows the radiatively efficient regime $L_{bol} \propto \dot{m}_{in}$ as long as $\dot{m}_{in} < 5$. Below this accretion rates, the SAD is indeed radiating all of its available energy. At $\dot{m}_{in} > 5$, the SAD enters the slim domain, where more and more energy becomes advected instead of being radiated away. As a consequence, the global luminosity has a steeper slope $L_{bol} \propto \dot{m}_{in}^{0.5}$.

As $r_J$ increases, the JED mode starts to have a bigger extent, and at any given accretion rate $\dot{m}_{in}$ the luminosity decreases as $r_J$ increases. This is the result of two different effects. First, as $r_J$ increases, more and more energy is controlled by the JED and transferred to the jets $b = P_{jets} / P_{acc, JED} = 0.3$ (papers I and II) instead of being radiated. Second, the JED thermal equilibrium is often strongly affected by advection, as $f = P_{adv} / P_{acc, JED} \propto \varepsilon^2$ is no more negligible (paper II). Combining these two effects, a more representative formulation is $L_{bol} \propto (1-b-f) ~ \dot{m}_{in}$, where $f \propto \varepsilon^2 = \varepsilon^2 (\dot{m}_{in})$ also is a function of accretion rate.

At low accretion rates $\dot{m}_{in} < 0.5$, the JED is optically thin and geometrically thick with $\varepsilon \simeq 0.2-0.3$ (termed thick disk solution in paper II). In the thick disk branch, the low density of the plasma allows $q_{ie} \propto n_e^{2}$ to be negligible. Ions are neither cooled down by radiation nor by Coulomb interactions: $T_i \gg T_e$. Contrarily to usual one-temperature plasmas, the disk thickness is then only linked to the ion pressure, $P_{gas,i} \gg P_{gas,e} + P_{rad}$, leading to $q_{adv} \simeq q_{adv,i} \gg q_{adv,e}$ (see Eq. (13) in paper II). In the ion thermal equilibrium from Eq. (\ref{eq:IONS}), advection is directly determined by ion heating, $q_{adv} \simeq q_{adv,i} \simeq (1-\delta) q_{turb}$, leading to $f = q_{adv} / q_{turb} \simeq 0.5$. Since $q_{adv,e} \ll q_{adv,i} \simeq (1-\delta) q_{turb}$, we obtain $q_{adv,e} \ll \delta q_{turb}$. In the electron equilibrium Eq. (\ref{eq:ELECTRONS}), radiation is then determined by $q_{rad} \sim \delta q_{turb} \propto \dot{m}_{in}$, leading to the trend $L_{bol} \propto \dot{m}_{in}$, unexpected in a thick disk\footnote{This in only true if $(1-\delta) \sim \delta$, which is the case here. In ADAFs, where $\delta = 1/2000$, even if $q_{adv,e} \ll (1-\delta) q_{turb}$ the factor $\delta \ll 1$ would not ensure that $q_{adv,e} \ll \delta q_{turb}$, and this reasoning would not stand.}. In the end, combining the loss of power through jets ($b$) and in advection ($f$) reduces the JED luminosity. For instance, for $r_J = 50$ (yellow color) it is reduced by a factor $1 / (1-b-f) \sim 5$ compared to the SAD mode power.

At high accretion rates $\dot{m}_{in} > 2$, the JED mode is optically thick and geometrically slim with $\varepsilon \simeq 0.1$ (termed slim disk solution in paper II). The bolometric luminosity now has a slope $L_{bol} \propto \dot{m}_{in}^{0.5}$ (see the previous discussion on the slim SAD mode), with a factor $\sim 2$ times lower in luminosity. Indeed, in this case, $f \simeq 0.15$, due to the lower temperature (and then the lower $\varepsilon$) compared to the thick disk solution.

Between these two solutions, from the thick to the slim disks, $f$ decreases from $0.45$ to $0.15$. The disk radiates more and more energy and the luminosity increases until it reaches the $L_{bol} \propto \dot{m}_{in}^{0.5}$ slope. During this transition, the slope is more abrupt and fits well with $L_{bol} \propto \dot{m}_{in}^{1.5}$. \\

The bottom panel of Fig.~\ref{fig:Lxmdot} shows that the $L_{3-9}$ variation with the accretion rate is different from $L_{bol}$. In the SAD mode ($r_J = r_{in} = 2$), $L_{3-9} \propto \dot{m}_{in}^{\sim 2}$ for $\dot{m}_{in} < 1$ and it slowly drops down to $L_{3-9} \propto \dot{m}_{in}^{0.5}$ while approaching the slim region. The JED mode with $r_J = 50$ remains closer to the bolometric behavior, with still $L_{3-9} \propto \dot{m}_{in}$ when $\dot{m}_{in} < 0.5$ and roughly $L_{3-9} \propto \dot{m}_{in}^{1.5}$ during the transition and in the slim region. At very low accretion rates, the JED mode radiates more energy in the $3-9$~keV band up to a factor $\sim 4$, while at higher accretion rates the SAD can radiate as high as $\sim 17$ times more energy in this range. \\

Two interesting comments can be done from Fig.~\ref{fig:Lxmdot}. First a "radiatively efficient accretion flow" does not necessarily mean $L \propto \dot{m}_{in}$. Indeed, as shown before in the JED dominated mode ($r_J = 50$) we find $L \propto \dot{m}_{in}$, while the JED radiates only few tens of its total available energy. The term "radiatively efficient" seems therefore inappropriate. Second, the evolution of the luminosity with the accretion rate strongly depends on the energy range used. In the SAD mode, while the disk is indeed radiatively efficient and $L_{bol} \propto \dot{m}_{in}$, the $3-9$~keV luminosity follows a $L_{3-9} \propto \dot{m}_{in}^{1.5-2}$ regime, which could be considered as the signature of a radiatively inefficient flow. This clearly means that the interpreation of the luminosity variation with the accretion rate in terms of radiative efficiency can be strongly misleading and should be done with caution.\\


Finally, Fig.~\ref{fig:Lxmdot} also illustrates that the functional dependence $L_{3-9} (\dot{m}_{in})$ can be much more complex than a single power law. This is quite promising as it is known that $L_{3-9} (\dot{m}_{in})$ may need to vary from one object to another \citep[][section~4.3.3 and Fig.~7]{Coriat11}. However, we cannot go further without considering a proper outbursting cycle. For instance, Fig.~\ref{fig:EffectOfRj} clearly shows that, in order to successfully reproduce the 2010-2011 cycle of GX 339-4 (black lines in Fig~\ref{fig:DFLD_complet}), one would need to (1) rise up, namely increase $\dot{m}_{in}$ from the quiescent state until the highest hard state; then (2) transit left by decreasing the transition radius $r_J$ until the full disappearance of the JED; (3) drop down in the soft realm by decreasing $\dot{m}_{in}$ and finally (4) transit right back to the hard zone by increasing $r_J$. It can thus be inferred from Fig.~\ref{fig:Lxmdot} that the evolution in time of the X-ray luminosity is sharper than $L_{3-9} \propto \dot{m}_{in}$ at high luminosities, because of the necessary decrease of $r_J$. A detailed modelling of the actual track followed by GX 339-4 during a full cycle will be presented in a forthcoming paper.

\subsection{Radio fluxes} \label{sec:RadioFluxes}

\begin{figure}[h!]
  \centering
  \includegraphics[width=\columnwidth]{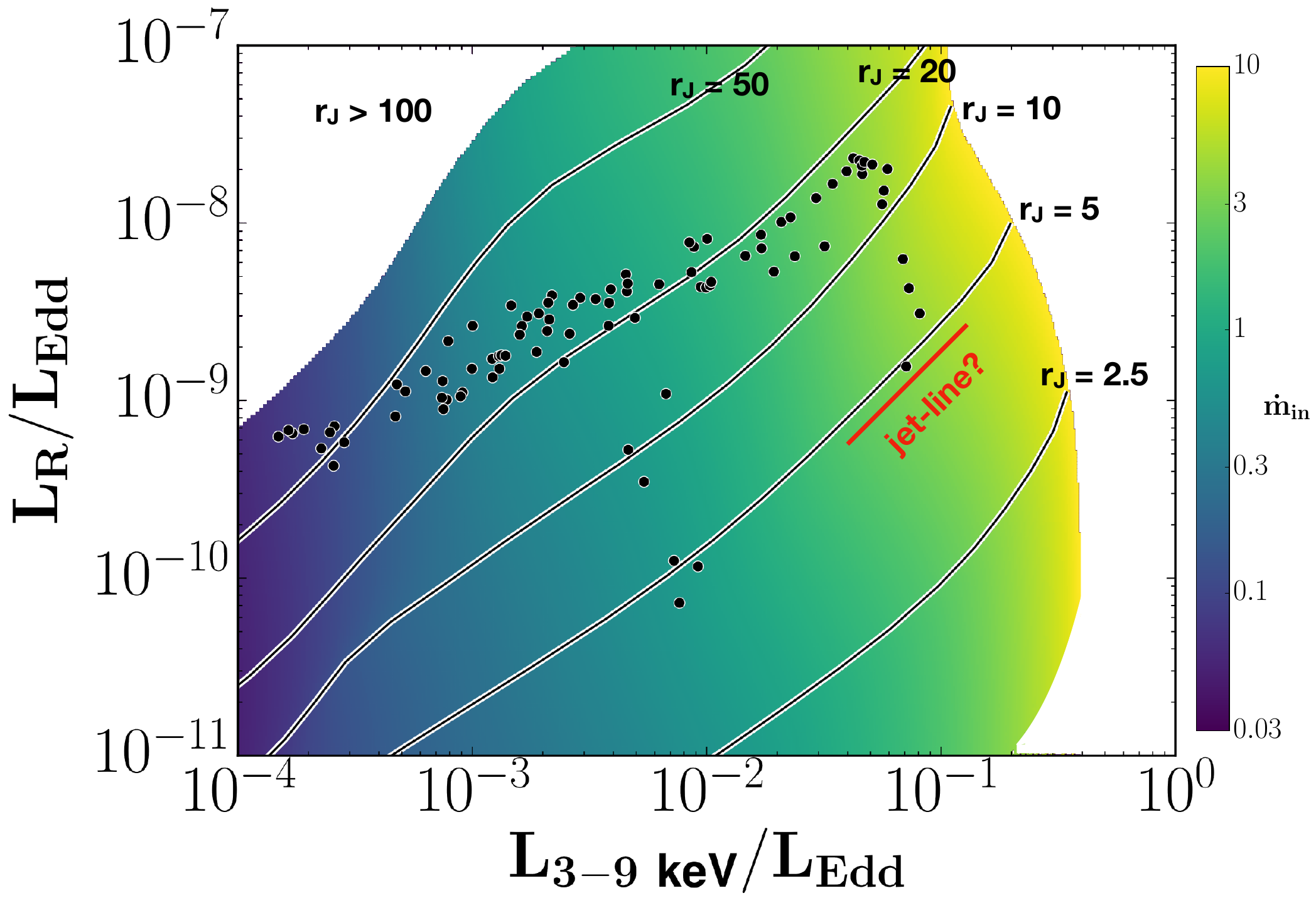}
   \caption{Power in the 8.6 GHz radio band in function of the 3-9 keV X-ray power (both in Eddington luminosity). Black solid lines are for constant transition radius $r_J$, while the color background shows the accretion rate $\dot{m}_{in}$. This figure has been done using a similar procedure than in Fig.~\ref{fig:EffectOfRj}. The black points are the observed values for GX 339-4 during its cycles between 2003 and 2011. A possible theoretical jet line has been drawn in red (see section~\ref{sec:RadioFluxes}).}
     \label{fig:PJET}
\end{figure}
%

For any given hybrid JED-SAD disk configuration, computed with a pair of parameters ($\dot{m}_{in}, r_J$), one can also derive an estimation of the one-sided jet radiative power $P_R$ emitted at the radio frequency $\nu_R = 8.6$~GHz. Assuming an electron distribution with $p=2$, Eq.~(\ref{eq:Fr}) leads to
\begin{equation}
P_R \equiv \nu_R L_R = \tilde f_R \, \dot{m}_{in}^{17/12} r_J \left ( r_J - r_{in} \right )^{5/6} \, L_{Edd}
\label{eq:PR}
\end{equation}
where the dimensionless factor $\tilde f_R$ incorporates factors such as $m$ or $r_{in}$, see Appendix \ref{sec:JetRadio}. For the sake of simplicity, we assume that this factor is a constant throughout the whole cycle. This is far from being obvious, but our goal here is simply to show that hybrid configurations have the potential to reproduce simultaneously both X-rays and jet radio emission. 
To have an estimate of $\tilde f_R$, we require that two radio observations at $\nu_R= 8.6$~GHz, one during the quiescent state and the other during the high-luminosity hard state (see their definition below), are qualitatively reproduced. We obtain $\tilde f_R = 3 \times 10^{-11}$. This allows us to compute the radio power $P_R$ as a function of ($\dot{m}_{in}, r_J$) and finally relate the radio power to the observed $3-9$~keV power.
In Fig.~\ref{fig:PJET}, the same procedure than in Fig.~\ref{fig:EffectOfRj} is used, but with binning of the true integrated luminosity $L_{3-9~\text{keV}}$ and radio power $P_R$. In addition, radio and X-ray observations of GX 339-4 were overplotted in black dots. It can be seen that most of them correspond to $r_J \sim 10-50$ while the accretion rate spans $\dot{m}_{in} \sim 0.01$ to almost $5$. Thus, reproducing the observed radio/X-ray diagram requires variations in mass accretion rate, from $\dot{m}_{in} < 0.1$ to almost $5$ here. However, in order to describe the disappearance or reappearance of the steady radio emission, i.e. the crossing of the jet line \citep{Fender04}, one needs to invoke variations in transition radius: a decrease in $r_J$ when the jet is quenched (bottom-right), and an increase when the jet re-appears (top-left). This is very promising for the model to reproduce observations, but further investigation needs to be done.

\begin{table*}[h!]
\centering
{\renewcommand{\arraystretch}{1.3} 
 \begin{tabular}{c | c c c c | c c | c c c c c}
 & \multicolumn{4}{c|}{Typical observed states} & \multicolumn{2}{c|}{Parameters} & \multicolumn{5}{c}{Results of fits} \\ \hline
Spectral & $\mathbf{PLf}$ & $\textbf{X-rays}$ & $\mathbf{\Gamma}$ & $\textbf{Radio}$ & $\mathbf{\dot{m}_{in}}$ & $\mathbf{r_J} $  & $\mathbf{PLf}$ & $\textbf{X-rays}$ & $\mathbf{\Gamma}$ & $\textbf{Radio}$ & $ \chi^2_{red}$\\
state & & $(\% L_{Edd})$ & & (mJy) & $(L_{Edd}/c^2)$ & $(R_g)$ & & $(\% L_{Edd})$ & & (mJy) & \\ \hline
\textbf{Q} & $ 1 $ & $ < 0.1 $ & $ 1.5-2.1 $ & $ 1 $& $ 0.06 $ & $ 100 $ & $ 1 $ & $ 0.11 \pm 0.01 $ & $ 1.9^{+0.3}_{-0.3}$ & $ 2.8 $ & $0.96$ \\
\textbf{LH} & $ 1 $ & $ 1 $ & $ 1.5-1.6 $ & $ 5 $ & $ 0.4 $ & $ 50 $ & $ 1 $ & $ 0.92 \pm 0.05 $ & $ 1.64^{+0.04}_{-0.04} $ & $ 11.5 $ & $1.10$ \\
\textbf{HH} & $ 1 $ & $ >10 $ & $ 1.6-1.8 $ & $ 25 $& $ 2 $ & $ 15 $ & $ 0.98 \pm 0.02 $ & $ 13.1 \pm 0.1 $ & $ 1.68^{+0.03}_{-0.02} $ & $ 20.2 $ & $1.60$ \\ 
\textbf{HS} & $ < 0.3 $ & $ 5 $ & $ 2-3 $ & $ < 0.01 $& $ 0.75 $ & $ 2 $ & $ 0.13 \pm 0.05$ & $ 4.6 \pm 0.1 $ & $ 2.4^{+0.3}_{-0.7} $ & $ 0 $ & $0.90$ \\
\textbf{LS} & $ < 0.3 $ & $ 1 $ & $ 2-3 $ & $ < 0.01 $& $ 0.45 $ & $ 2 $ & $ 0.2 \pm 0.05 $ & $ 1.40 \pm 0.04 $ & $ 2.5^{+0.3}_{-0.6} $ & $ 0 $ & $1.04$ \\ \hline
\end{tabular}}
 \caption{Typical observed properties of the five canonical states (left), pairs of parameters ($\dot{m}_{in}$ and $r_J$, center) and XSPEC fits (right) results associated to the five chosen canonical states Q, LH, HH, HS and LS. Fits are performed using the simple model described in section~\ref{sec:Py2Xspec}, and the definition of the five states is detailed in section~\ref{sec:XrB5states}.}
 \label{table:States}
\end{table*}

\subsection{XrB canonical spectral states} \label{sec:XrB5states}

\begin{figure}[h!]
\centering
\sidecaption
  \includegraphics[width=1.\columnwidth]{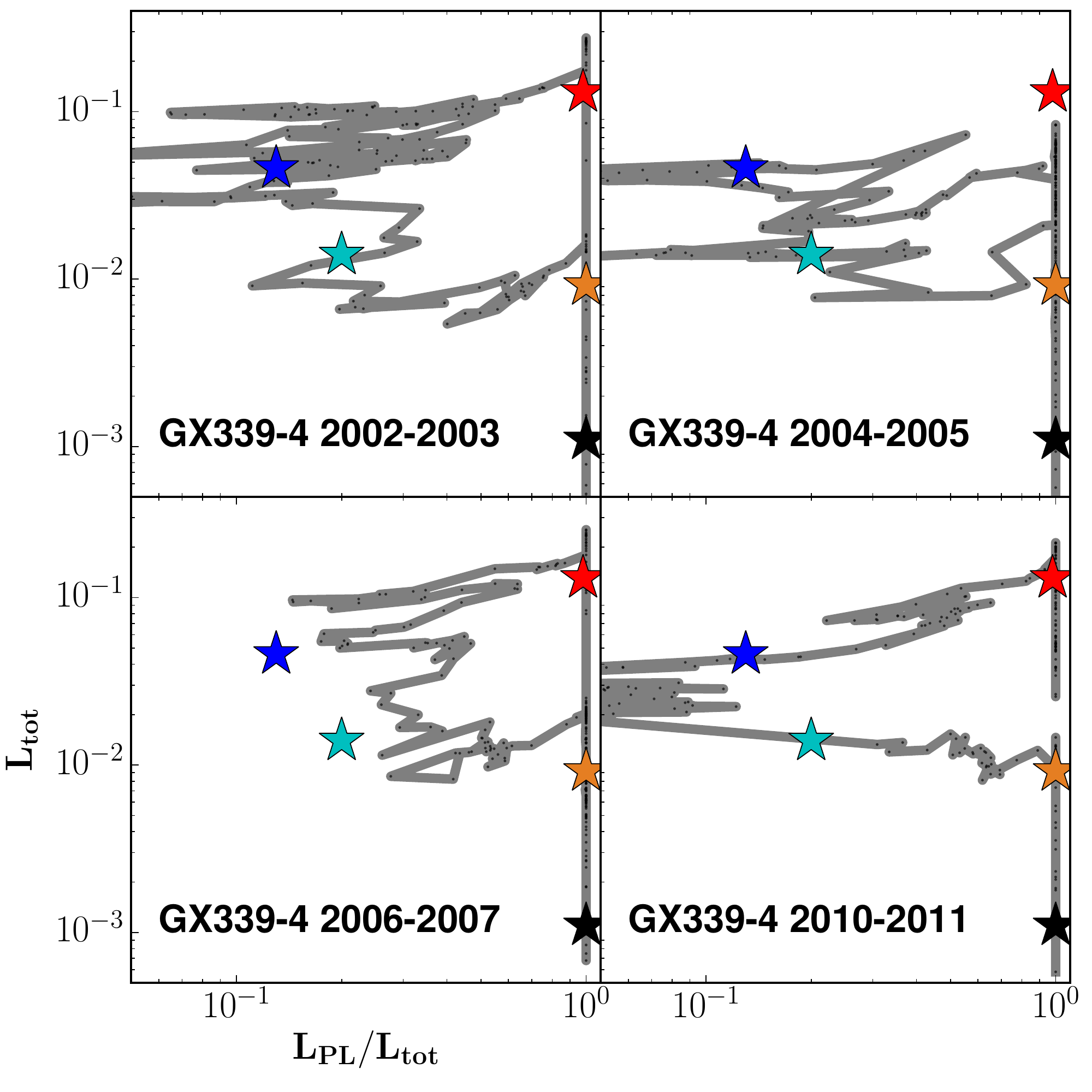}
   \caption{DFLDs for the past outbursts of GX 339-4 between MJD$50290$ and MJD$55650$ extrapolated in the $3-200$~keV energy range\footref{fnt:ranges}. A typical cycle goes from Q (black), and crosses LH (orange) up to HH (red). It then transits to the HS (blue), decreases to LS (cyan) until it transits back to LH, before decreasing down to Q. The stars mark the positions of the five canonical spectral states Q, LH, HH, HS and LS defined in section~\ref{sec:XrB5states}.}
     \label{fig:DFLD_complet}
\end{figure}

In section \ref{sec:EffectOfRjMdot}, it is shown that hybrid JED-SAD configurations can cover the observed DFLDs by varying independently $\dot{m}_{in}$ and $ r_J$. 
We focus here on the five typical spectral states that any given XrB needs to cross (or get close to) when making a full cycle, and detail their characteristics in our JED-SAD framework. These five states are shown in Fig.~\ref{fig:DFLD_complet} and are named quiescent state (hereafter Q), low-luminosity hard state (LH) and low-luminosity soft state (LS), both at the soft-to-hard lower transition branch, and high-luminosity hard state (HH) and high-luminosity soft state (HS) both at the hard-to-soft upper transition branch.

\begin{figure*}[h!]
 \centering
 \includegraphics[width=1.\textwidth]{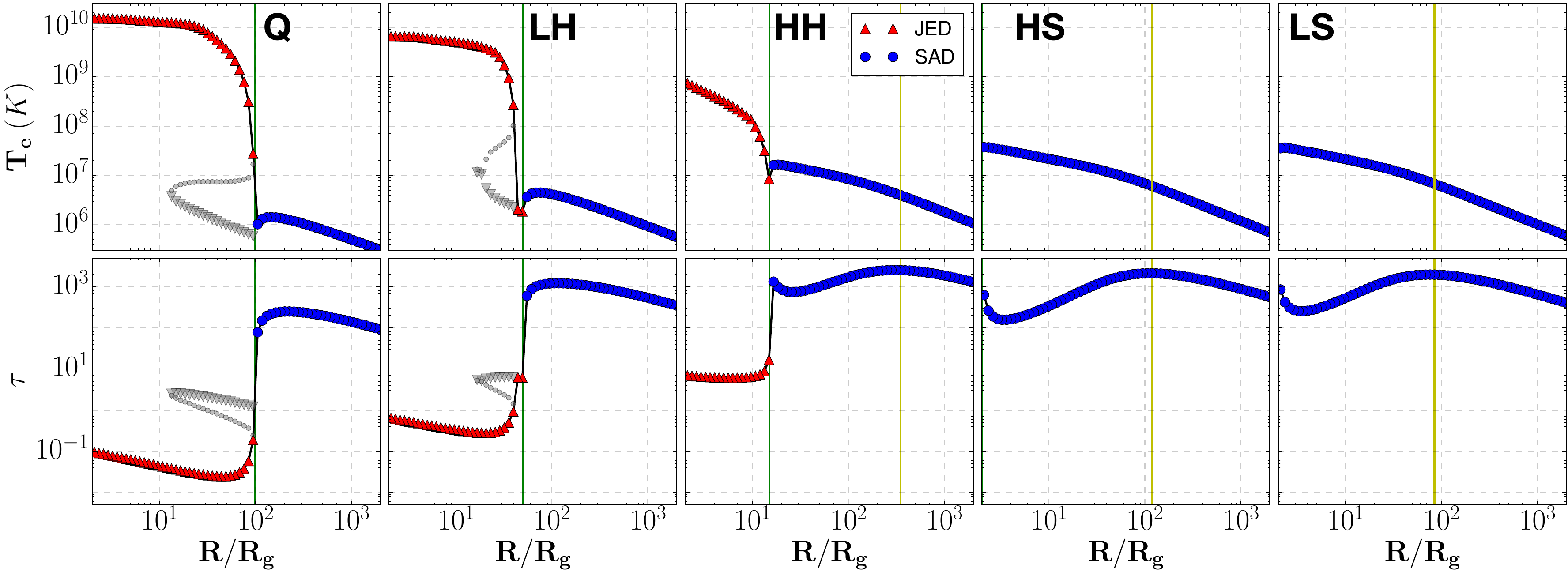}
  \caption{Computed radial structures of hybrid JED-SAD disk configurations associated to the canonical states defined in Table~\ref{table:States}. From left to right: \textbf{Q} for quiescent state, \textbf{LH} for low-hard, \textbf{HH} for high-hard, \textbf{HS} for high-soft and \textbf{LS} for low-soft. Top: electron temperature $T_e$ at the disk mid plane. Bottom: Thomson optical depth $\tau_T$. Red triangles show the JED zone, and blue dots describe the SAD zone. The vertical green line marks their separation at the transition radius $r_J$. The vertical yellow line marks the transition from a gas to radiation pressure supported regime within the SAD. In addition, the other two possible thermal solutions are shown in grey when present, the unstable in circles and the thin disk in triangles, see paper II.}
  \label{fig:TypicalStates}
\end{figure*}

The quiescent state chosen is clearly not the most quiescent state that can be reached by an XrB. It is at position in the DFLD that any object needs to cross while going up and down. Our choice of the two soft states in the DFLD is somewhat arbitrary as it depends on the chosen level of the hard tail, see section~\ref{sec:HardTail}. Our computed spectra are also shown from $0.5$ to $500$~keV but remember that the observational PCA and HEXTE data are available only from $3$ to $200$~keV. \\

\begin{figure*}[h!]
 \centering
 \includegraphics[width=1.\textwidth]{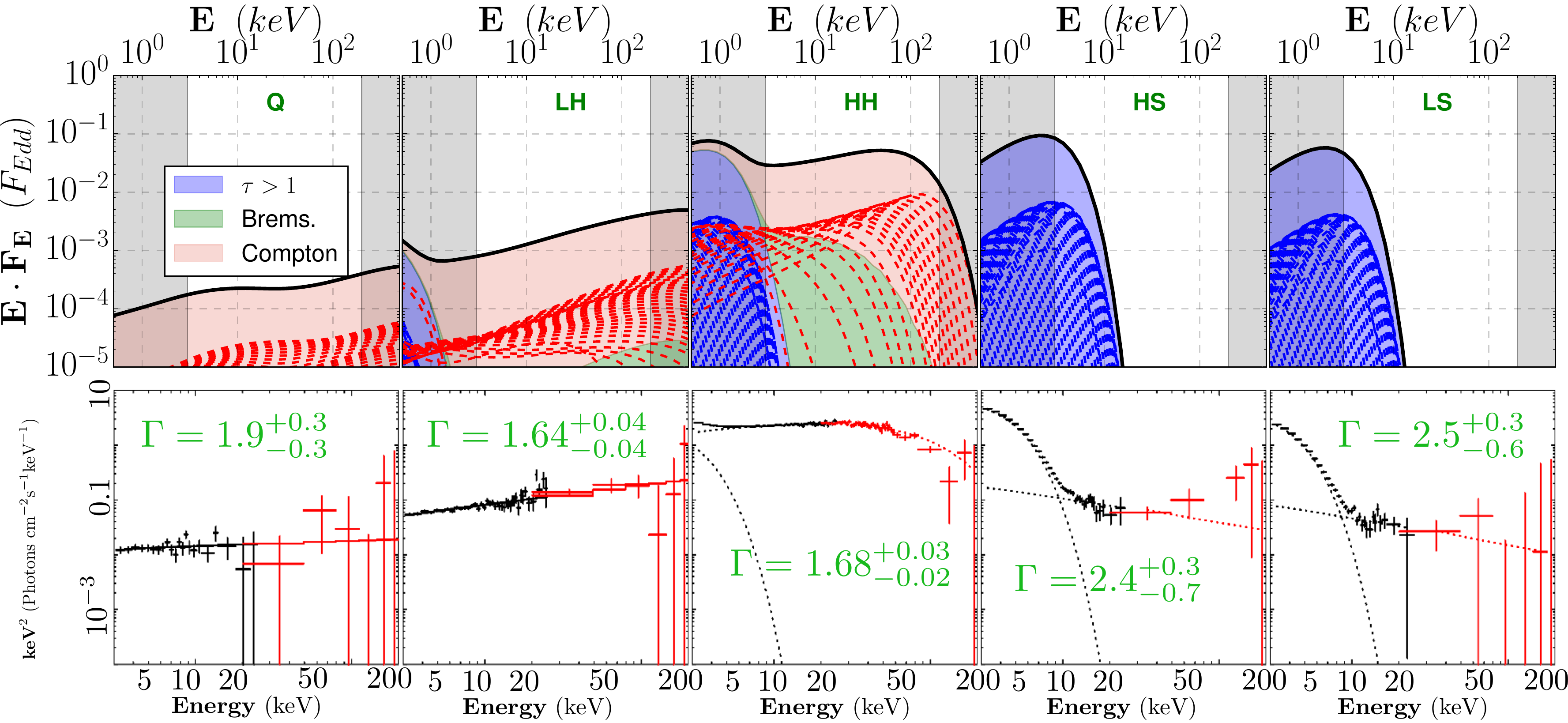}
  \caption{Theoretical SED (top) and XSPEC spectral fits (bottom) for the five canonical spectral states computed in Fig.~\ref{fig:TypicalStates}. It can be seen that hard states spectra are always dominated by the comptonization of soft photons, mostly due to local Bremsstrahlung and cold photons from the outer SAD. The white area in the theoretical SED corresponds to the observationally relevant $3-200$~keV energy band. The value of the spectral index $\Gamma$ is shown for each state, with its errors.}
  \label{fig:SpectralStates}
\end{figure*}
\begin{figure}[h!]
\centering
  \includegraphics[width=1.\columnwidth]{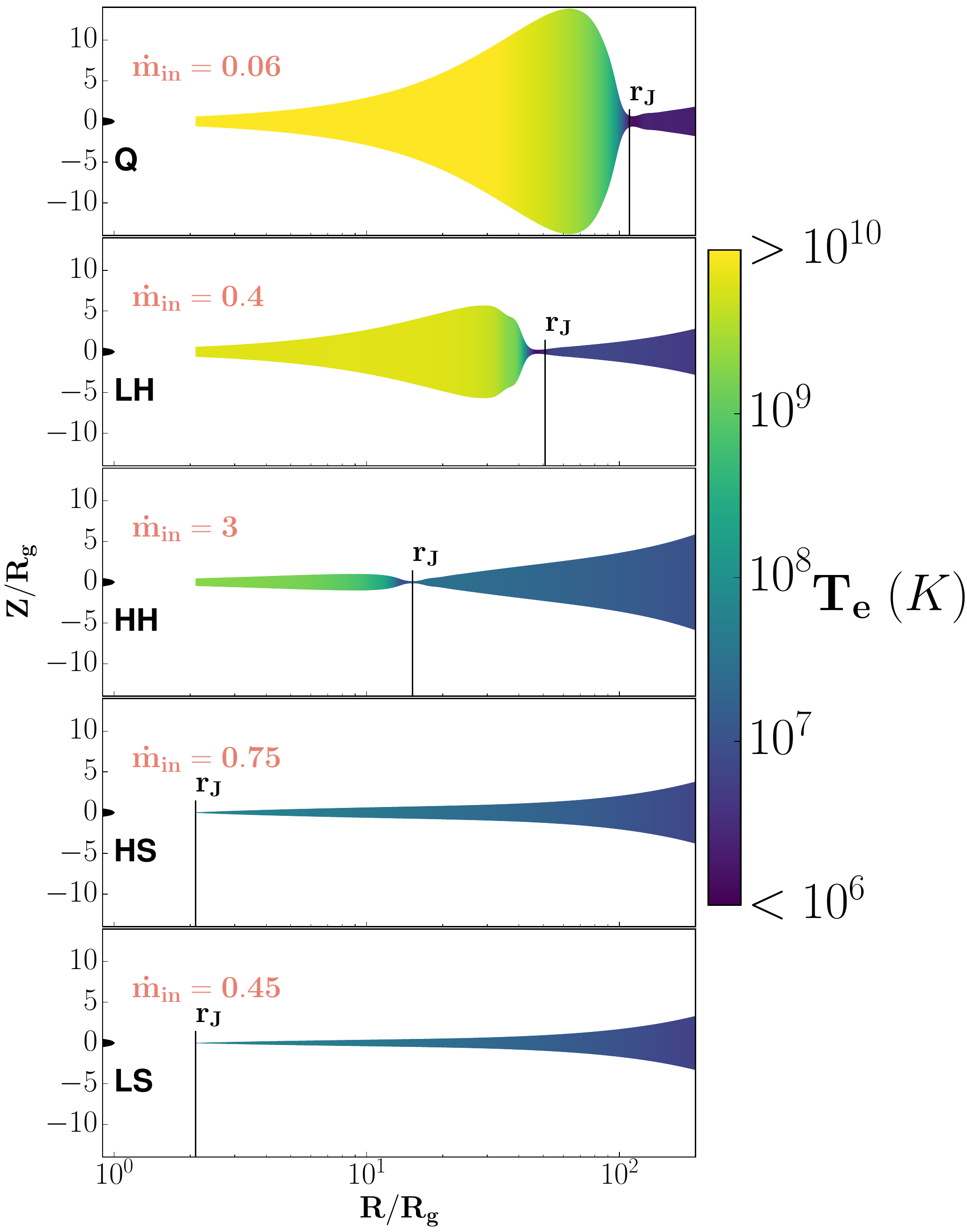}
   \caption{Computed geometrical shape of the hybrid disk, consistent with the dynamical resolution (Fig.~\ref{fig:TypicalStates}) and SED (Fig.~\ref{fig:SpectralStates}) for each of the five canonical states. The color background is the central electron temperature. Note that the X-scale is logarithmic and the Y-scale is linear.}
     \label{fig:JoliesImages}
\end{figure}

Table~\ref{table:States} middle panel shows the values of the parameters $\dot{m}_{in}$ and $r_J$ that better characterize these five canonical states. We also have reported in this table the values (and their associated $3 \sigma$ errors) of the power-law fraction $PLf$, X-ray luminosity and spectral index $\Gamma$ derived from XSPEC fits, as well as the expected radio flux density at $8.6$~GHz, $F_{8.6~\text{GHz}}$. These fluxes have been computed in mJy using Eq.~(\ref{eq:PR}), namely $F_{8.6~\text{GHz}}= 10^{26} \times P_R/(4\pi D^2 \nu_R)$, with $\nu_R= 8.6$~GHz and $\tilde{f}_R = 3 \times 10^{-11}$. Figs.~\ref{fig:TypicalStates} and \ref{fig:SpectralStates} illustrate the thermal state radial distribution (temperature, optical depth) and the theoretical and faked spectra. An accurate representation of the physical structure (size) and temperature (color) of the disk in those five states is shown in Fig.~\ref{fig:JoliesImages}.
 
In the following, we discuss each of the canonical states with more details. \\

\noindent {\bf - Q state:} the quiescent state has an X-ray luminosity lower than $0.1 \%$ Eddington, with a typical power-law spectrum of index $\Gamma \simeq 1.5-2.1$ in the observed $3-200$~keV band \citep{Remillard06}. It exhibits faint but steady radio and IR luminosities fluxes \citep{Fender01, Corbel13}, probing weak but detectable jets. 
It is located at the bottom-right of the DFLD. In our JED-SAD framework it is characterized by a very low accretion rate $\dot{m}_{in} \lesssim 1$ and a relatively high transition radius $r_J \gg r_{in}$. In the example shown in this section we choose $\dot{m}_{in} = 0.06$ and $r_J = 100$. The innermost region of the disk (from $r \simeq 30$ down to $r_{in}$) is optically thin with electron temperature as high as $T_e \simeq 10^{10}$~K (Fig.~\ref{fig:TypicalStates}, left panel). This results in a global spectrum that is the sum of multiple power-law spectra with roughly the same shape $\Gamma \sim 1.6-2$ and $E_{cut} \gg 200$~keV (Fig.~\ref{fig:SpectralStates}, left panel), in good agreement with the observations. In addition, the power available in the jets is relatively small due to a very low accretion rate (Table \ref{table:States}).\\


\noindent {\bf - LH state:} the low-luminosity hard state is characterized by a power-law dominated spectrum, with spectral index $\Gamma \simeq 1.5-1.6$, no cutoff detected $E_{cut} > 200$~keV \citep[][and references therein]{Grove98, Zdz04} and a typical luminosity $L_{tot} \simeq 1 \% L_{Edd}$. This state is also associated to steady and high radio and IR luminosities suggesting powerful jets. In this article we choose $\dot{m}_{in} = 0.4$ and $r_J =50$. As shown in Fig.~\ref{fig:TypicalStates} top panel, the temperature of this flow increases quickly from $T_e \simeq 10^{6}$~K in the outer parts of the JED to $T_e \gtrsim 5 \times 10^{9}$~K in the inner parts for most of the JED extension (from $r = 20$ down to $r_{in}$). This state does not strongly depend on $r_J$ as long as it is larger than a few tens of $r_{in}$, as the global spectrum is the sum of similar spectra with $\Gamma \sim 1.2-1.8$ (Fig.~\ref{fig:SpectralStates}, LH-panel). This configuration is also accompanied by more powerful jets, due to larger accretion rate compared to Q-states, in agreement with stronger observed radio emission. \\

\noindent {\bf - HH state:} the high-luminosity hard state is also characterized by a power-law dominated spectrum, with a spectral index $\Gamma \simeq 1.6-1.8$, but with a high energy cutoff generally detected $E_{cut} \simeq 50-200$~keV \citep{Motta09} and luminosities as high as $30 \% L_{Edd}$ \citep{Dunn10}. Actually, as the luminosity increases, $E_{cut}$ is observed to decrease from $>200$~keV to $\sim 50$~keV, while the spectral index slightly changes from $\Gamma = 1.6$ to $\Gamma = 1.8$ before transiting to the soft state \citep[see Figure 6 in][]{Motta09}. Those states also show the highest radio and IR fluxes \citep{Coriat09}, suggesting the most powerful jets of the cycle. In our JED-SAD framework this state is characterized by a larger accretion rate $\dot{m}_{in} > 1$, we choose $\dot{m}_{in} = 3$. As shown in paper II, at such high accretion rate the hot geometrically thick disk solution switches to a denser and cooler solution, the so-called slim disk. The disk is optically slim $\tau \sim 1-10$ and rather warm $T_e \sim 10^{8-9}$~K (Fig.~\ref{fig:TypicalStates}, middle panel). The spectra associated to those slim disks solutions are closer to a very hot multi-temperature disk blackbody emission (paper II). The combination between electron temperature and optical thickness distribution with radius produces a spectral shape in agreement with observations (i.e. $\Gamma \simeq 1.6-1.8$ and $E_{cut} \in [50, ~ 200]$~keV). Once we choose $\dot{m}_{in} = 3$, the range of appropriate values for transition radius is rather narrow to reproduce the value of $\Gamma$, we adopt $r_J = 15$. The large $\dot{m}_{in}$ and $r_J$ result in a large jet power consistent with observations (see Table~\ref{table:States}).

\noindent {\bf - HS state:} the high-luminosity soft state is defined by a dominant multi-temperature blackbody with maximum effective temperature $T_{in} \lesssim 1$~keV and total flux $L_{tot} \sim 5 \% L_{Edd}$. In addition, many of the soft states display a minor component, the hard tail (see section \ref{sec:HardTail}). These states are also characterized by the absence of steady radio emission, interpreted as the jet disappearance \citep[][]{Fender99, Corbel00}. In our JED-SAD framework this translates to $r_J = r_{in}$, i.e. our accretion flow is entirely in a SAD mode. In the example shown in Fig.~\ref{fig:TypicalStates}, we choose $\dot{m}_{in} = 0.75$ with a $10\%$ hard tail. \\

\noindent {\bf - LS state:} the spectral shape of the low-luminosity soft state is similar to the HS, with a typical maximum effective disk temperature $T_{in} \gtrsim 0.6-0.7$~keV, a $\Gamma \simeq 2-3$ hard tail, and total luminosity $\sim 5$ times lower. 
We define our canonical LS state with the same level of hard tail. In our JED-SAD framework, this correspond to lower $\dot{m}_{in}$ but still $r_J = r_{in}$ (no JED). In the example shown in this section, we choose $\dot{m}_{in} = 0.45$. The absence of JED means there are no jets, i.e. no radio emitted. \\

\section{Concluding remarks} 
\label{sec:Ccl}

In paper II, we studied the thermal equilibrium of jet-emitting disks (JED). JEDs are assumed to be thread by a large scale vertical magnetic fields, building two jets that produce a torque responsible for supersonic accretion. 

In this article, we extend the code to compute thermal equilibria of hybrid disk configurations. This configuration assumes an inner JED and an outer standard accretion disk (SAD), characterized by a highly subsonic accretion speed. The transition between those two flows is assumed to be abrupt ($\Delta R / R \ll 1$) at some transition radius $r_J$. As argued in section~\ref{sec:inter}, such a transition requires a discontinuity in the disk magnetization $\mu$ that can be obtained if the transition radius $r_J$ is a steep density front. The transition radius $r_J$ would thus correspond to a density front advancing or receding within the disk during an outburst, as also found in the context of ADAF-SAD transitions \citep{Honma96, Manmoto00}. Why such a density front would be present is an open question, possibly answered by how matter is initially brought in towards the disk inner regions. In any case, if such a front is indeed produced, it is not clear how it would be maintained over the long duration of the outburst.

Now, such a density front is known to be favorable to the Rossby wave instability \citep[][and references therein]{Tagger99, Lovelace99, Li00, Tagger04, Meheut10}, which leads to the formation of non-axisymmetric vortices within the disk. Whether or not the density front is smeared out and destroyed or simply perturbed (leading possibly to quasi-periodic oscillations) remains to be investigated. We refer the interested reader to the discussion on timing properties in paper I, section 4.

On the other hand, one might argue as well that such a discontinuity in the disk magnetization is unrealistic and that, instead, there is a continuous increase in $\mu$ towards the disk inner regions \citep{Petrucci08}. Assuming that such a situation would be indeed realized, the transition radius $r_J$ required in our spectral calculations would then be interpreted as the transition from the outer optically thick disk to the inner optically thin disk. Correspondingly, one could argue that the outer low-magnetized disk regions would give rise to winds whereas jets would be launched from the inner highly magnetized disk regions (JED). The difficulty with this scenario is that it relies on the disk mass loss and the radial distribution of the large scale vertical magnetic field, both unknown to date. Our simple approach, that assumes a sharp JED-SAD radial transition, can be seen as a first step towards addressing this difficult topic in XrB accretion disks. \\

The outside-in radial transition in accretion speed translates thermally, from an outer optically thick and cold accretion flow to an inner optically thin and hot flow. The soft photons emitted by the outer disk also provide a non-local cooling term which, added to advection of internal energy, allow a smooth thermal transition between these two regions. For a given JED-SAD dynamical solution the corresponding spectrum depends only on the mass accretion rate onto the black hole $\dot{m}_{in}$ and the transition radius $r_J$ between the two flows. We explore in this article a large range in $\dot{m}_{in}$ and $r_J$. Using XSPEC, we build synthetic spectra and fit them using a standard observers procedure (section~\ref{sec:Py2Xspec}), allowing us to easily compare the resulting fits to observations.

We show that this framework is able to cover the whole domain explored by typical cycles in a disk fraction luminosity diagram (Fig.~\ref{fig:EffectOfRj}). Furthermore, five canonical X-ray spectral states representative of a standard outburst are quantitatively reproduced with a reasonable set of parameters (Figs.~\ref{fig:TypicalStates} to \ref{fig:JoliesImages} and Table~\ref{table:States}). A very interesting and important aspect of this framework is its ability to simultaneously explain both X-ray and radio emissions (Fig~\ref{fig:PJET} and Table~\ref{table:States}). In a forthcoming paper we will show the required time sequences $\dot{m}_{in}(t)$ and $r_J(t)$ needed to reproduce a full cycle within the JED-SAD paradigm. \\


\begin{acknowledgements}
   We are grateful to the anonymous referee for his/her careful reading of the manuscript. The authors acknowledge funding support from French Research National Agency (CHAOS project ANR-12-BS05-0009 http://www.chaos-project.fr), Centre National de l'Enseignement Superieur (CNES) and Programme National des Hautes Energies (PNHE) in France. SC is supported by the SERB National Postdoctoral Fellowship (File No. PDF/2017/000841).
\end{acknowledgements}

\newpage 

\bibliographystyle{aa} 
\bibliography{ADSbibnew.bib}

\appendix 

\section{Jet radio emission} \label{sec:JetRadio}

We follow here the same reasoning as \citet{BK79} and \citet{Heinz03}. Jets emit synchrotron radiation from a non-thermal power-law distribution satisfying $\frac{dn_e}{d\gamma} = C \gamma^{-p}$, where $\gamma$ is the particle Lorentz factor, $p$ is the power-law index and $C$ the normalization factor. This factor is related to the pressure of the relativistic particles and is usually assumed to follows the magnetic field pressure so that $C= C_o B^2$ where $C_o$ is a constant. 
The synchrotron self-absorption coefficient $\alpha_\nu$ and emissivity $j_\nu$ for a non-thermal particle distribution are \citep{RL79}
\begin{eqnarray}
\alpha_\nu &=& A_p C B^{\frac{p+2}{2}} \nu^{- \frac{p+4}{2}} \nonumber \\ 
j_\nu &=& J_p C B^{\frac{p+1}{2}} \nu^{- \frac{p-1}{2}}  
\end{eqnarray}
where $A_p$ and $J_p$ are proportionality constants and weakly dependent on $p$. At a given distance $Z$ from the source, the jet has a radius $R_{jet}(Z)$ and a finite width $\Delta R_{jet}(Z)$. This allows to compute the synchrotron optical depth to self-absorption $\tau_\nu = \Delta R_{jet} \alpha_\nu$. The local emitted spectrum is $I_\nu = S_\nu \left (1- e^{-\tau_\nu}\right )$ with a source function $S_\nu= j_\nu/\alpha_\nu$. At high frequencies in the optically thin regime, the spectrum $I_\nu \simeq \Delta R_{jet} j_\nu \propto \nu^{- \frac{p-1}{2}}$ is decreasing with the frequency whereas at low frequencies, in the optically thick regime, $I_\nu \simeq S_\nu \propto \nu^{5/2}$. At any given altitude, the self-absorbed spectrum is therefore peaked at a frequency $\nu_c$ defining the jet photosphere. It is such that $\tau_{\nu_c} = 1$, namely
\begin{equation}
\nu_c = \left (A_p C_o\right )^{\frac{2}{p+4}} \Delta R_{jet}^{\frac{2}{p+4}} B^{\frac{p+6}{p+4}}
\label{eq:nuc}
\end{equation}
and the jet surface brightness becomes
\begin{equation}
I_{\nu_c} =J_p A_p^{-\frac{p-1}{p+4}} C_o^{\frac{5}{p+4}} \Delta R_{jet}^{\frac{5}{p+4}} B^{\frac{2p+13}{p+4}} \nonumber
\end{equation}

The monochromatic flux received at a frequency $\nu_c$ from a one-sided jet of width $2R_{jet}$, viewed side-on at a distance $D$ is 
\begin{equation}
F_{\nu_c} = \int I_{\nu_c} d\Omega = \frac{2}{D^2} \int_{R_g}^\infty dZ R_{jet} I_{\nu_c} = F_o \int_1^\infty dz r_{jet}  \Delta r_{jet}^{\frac{5}{p+4}} b^{\frac{2p+13}{p+4}}
\label{eq:Fnu}
\end{equation}
where the distances have been normalized to the gravitational radius ($z=Z/R_g, r=R/R_g$), $b=B/B_{in}$ where $B_{in}$ is a fiducial magnetic field and 
\begin{equation} 
F_o = \frac{2}{D^2} J_p A_p^{-\frac{p-1}{p+4}} C_o^{\frac{5}{p+4}} R_g^{\frac{2p+13}{p+4}} B_{in}^{\frac{2p+13}{p+4}} \nonumber
\end{equation}
The amplitude of the fiducial magnetic field $B_{in}$ depends on the underlying jet model. In our case, we assume that it is the innermost (the largest) magnetic field within the JED, namely 
\begin{equation} 
B_{in} = B_z(r_{in}) =  \sqrt{\mu_o P_*} \left ( \frac{\mu}{m_s}\right )^{1/2}\dot{m}_{in}^{1/2} r_{in}^{-5/4} \nonumber
\end{equation}
where $\mu_o$ is the vacuum permittivity and $P_*=m_p c^2/\sigma_T R_g$ with $m_p$ the proton mass, $c$ the speed of light and $\sigma_T$ the Thomson cross section. 

At any given altitude $z$, the jet width $\Delta r_{jet}$ is proportional to the radial extent of the jet-emitting region. Within a self-similar ansatz, namely using the self-similar variable $x=z/r$, one would simply write $\Delta r_{jet}= (r_J - r_{in}) f_r(x)$, where $f_r(x)= R(Z)/R_o$ is the self-similar function providing the cylindrical radius $R(Z)$ of a field line anchored at a radius $R_o$ within the disk. Following the same idea, the function $b$ would be only a function of the self-similar variable $x$ and $r_{jet}= r_J f_r(x)$. Equation (\ref{eq:Fnu}) can then be written 
\begin{equation} 
F_{\nu_c} = F_o (r_J - r_{in})^{\frac{5}{p+4}} r_J r_{in}  \int_{1/r_{in}}^\infty dx \left ( f_r b \right)^{\frac{2p+13}{p+4}} \nonumber
\end{equation}
where we used $dz=r_{in} f_r(x)dx$. The integral only depends on the jet dynamics and thus, for a given frequency, the received flux scales as  
\begin{equation}
F_{\nu_c} \propto   r_{in}^{- \frac{6p+49}{4p+16}}  \, 
\left ( \frac{\mu \, m \, \dot{m}_{in}}{m_s} \right)^{\frac{2p+13}{2p+8}} \,  r_J^{\frac{p+9}{p+4}}\,  \left (1- \frac{r_{in}}{r_J} \right )^{\frac{5}{p+4}} \nonumber
\end{equation}
allowing to compute the monochromatic power $L_\nu=4\pi D^2 F_\nu$ emitted by an XrB at a distance $D$. For a JED with $\mu/m_s \sim 1$, this leads to Eq.~(\ref{eq:Fr}) with $p=2$ and all proportionality constants incorporated in the dimensionless coefficient $f_R$. Note that this result has been obtained using an exact self-similar ansatz but it would remain valid as long as jets from black hole systems obey a more general similarity law (see discussion in \cite{Heinz03}). \\


This simple model also allows to derive the jet spectrum in the optically thick regime under quite generic assumptions. Equation (\ref{eq:nuc}) shows that for a given frequency $\nu_c$, there will be a distance $Z_c$ associated. Parts of the jet below and above $Z_c$ will provide a negligible contribution to the overall spectrum at that frequency. As the jet width varies like $\Delta R_{jet}(Z) \propto R_{jet} (Z)$ and assuming $B \propto R_{jet}^{-\delta}$ as well as 
$R_{jet} \propto Z^{1/\omega}$, one gets after some algebra $F_{\nu_c} \propto \nu_c^{\alpha_p} $ where the power law index is
\begin{equation}
\alpha_p = \frac{(2p+13)\delta - (p+4)\omega - (p+9)}{(p+6)\delta - 2} 
\label{eq:alpha}
\end{equation}
Although one should not pay too much attention to this simple expression, it allows nevertheless to grasp interesting relations between the jet spectrum and the underlying physics. 

The index $\omega$ describes the degree of collimation of the jet. Collimated flows require $\omega$ to be larger than unity (cone), a value of 2 (parabole) or slightly larger being acceptable. The index $\delta$ is more complex as it depends on the dominant magnetic field in the jet region. If $B \sim B_z >> B_\phi$, then a value $\delta = 2$ would correspond to magnetic flux conservation in a jet with almost no toroidal magnetic field (hence no electric current). If, on the contrary, $B \sim B_\phi >> B_z$, then a value $\delta = 1$ is more likely as it describes the existence of a constant asymptotic current (hence collimation). We therefore expect $1 \leq \delta \leq 2$. Now, Eq.~(\ref{eq:alpha}) shows that for any $\omega < \omega_o = ((2p+13)\delta - (p+9) )/(p+4)$, one gets $\alpha_p>0$. This shows that the less collimated the jet, the steeper the spectrum. 

For the conventional value $p=2$, $\omega_o= (17\delta - 11)/6$ and is always larger than unity for $\delta >1$. Flat spectrum sources with $\alpha_2=0$ could then be described by a jet structure such that $\delta= (11+6\omega)/17$. 
This could be realized for instance with $\omega=1$ and $\delta=1$. However the presence of an asymptotic current is inconsistent with a conical jet shape \citep{1989ApJ...347.1055H} so it can be ruled out. The other extreme possibility, with $B \sim B_Z$ or $\delta=2$, leads to $\omega=23/6=3.83$, a highly collimated flow which requires usually the presence of an important axial current (thus some $B_\phi$). A more reasonable jet profile, such as $\omega=2$ (paraboloidal jet), requires $ \delta = 23/17 = 1.35$, a profile expected in a helical jet structure. This is perfectly reasonable and advocates for self-confined magnetized jets as the source of the observed flat spectra in radio bands. \\

The monochromatic jet power is $L_\nu = 4\pi D^2 F_\nu \propto \nu^{\alpha_p}$. If one considers that the jet power-law spectrum is established within a range $[\nu_{min},\nu_{max}]$, then the total bolometric jet power is
\begin{equation}
L_{bol} = \nu_{max} L_{\nu_{max}} \frac{1- \left ( \frac{\nu_{min}}{\nu_{max}} \right )^{1+\alpha_p}}{1+\alpha_p} \nonumber
\end{equation}
For a flat spectrum source ($\alpha_p \simeq 0$), the radiative losses are dominated by the highest frequency $\nu_{max}>> \nu_{min}$. This translates into the convenient expression 
$L_{bol} \simeq \nu_{max} L_{\nu_{max}} = \nu_R L_R (\nu_{max}/\nu_R)^{1+\alpha_p}$, where $L_R$ is the monochromatic power emitted at the radio frequency $\nu_R$. The maximum frequency is observationally determined as the break frequency $\nu_B$. Making use of Eq.~(\ref{eq:Fr}), allows to derive the radiative efficiency for a (one-sided) jet
\begin{equation}
f_{rad} = \frac{L_{bol}}{P_{jet}} = \frac{4}{b} f_R \left ( \frac{\nu_B}{\nu_R} \right)^{1+\alpha_p} (m \dot{m}_{in})^{ \frac{5}{2p+8}}  
r_{in}^{- \frac{2p+33}{4p+16}} \, 
r_J^{\frac{p+9}{p+4}}\,  \frac{ \left (1- \frac{r_{in}}{r_J} \right )^{\frac{5}{p+4}}}{1- \left ( \frac{r_J}{r_{in}} \right )^{\xi-1} } \nonumber
\end{equation}
which needs to be always (much) smaller than unity. It appears to be indeed the case if one uses the value $f_R \sim 1~ 10^{-10}$ (or $\tilde{f}_R \sim 1~ 10^{-10}$) derived in section~\ref{sec4} and taking $\nu_B \sim 10^{14}$ Hz, $\nu_R \sim 10^{10}$ Hz with $\alpha_2 =0$.

\end{document}